\documentclass[showpacs,showkeys,preprintnumbers,amsmath,amssymb]{revtex4}

\usepackage{graphicx}

\def\thebibliography#1{
 \list
 {[\arabic{enumi}]}{\settowidth\labelwidth{[#1]}\leftmargin\labelwidth
 \advance\leftmargin\labelsep
 \usecounter{enumi}}
 \def\newblock{\hskip .11em plus .33em minus -.07em}
 \sloppy
 \sfcode`\.=1000\relax}

\begin{document}

\markboth{F. C. G. A. Nicolleau \& J. C. Vassilicos}
{Wavelet analysis of Wave motion}


\title{WAVELET ANALYSIS OF WAVE MOTION}

\author{F. C. G. A. NICOLLEAU\footnote{
Corresponding author.}}

\address{SFMG, Department of Mechanical Engineering, The University of Sheffield,
\\
Mappin Street S1 3JD Sheffield, UK}

\author{J. C. VASSILICOS} 

\address{Department of Aeronautics,
Imperial College of Science, Technology and Medicine,
\\
Prince Consort Road, South Kensington, London SW7 2BY, UK}


\begin{abstract}
In this paper
high resolution wave probe records
are examined using
wavelet techniques with a view to determining the sources and relative
contributions of capillary wave energy along representative wind wave forms.
Wavelets enable computations of conditional spectra and turn out to be
powerful tools for the study of the development and propagation of capillary
waves. They also enable the detailed analyses of the relative
contributions to the spectrum of the
wave peaks and troughs.
\end{abstract}

\maketitle

\section{Introduction}

\begin{figure}
\includegraphics[width=12.cm]{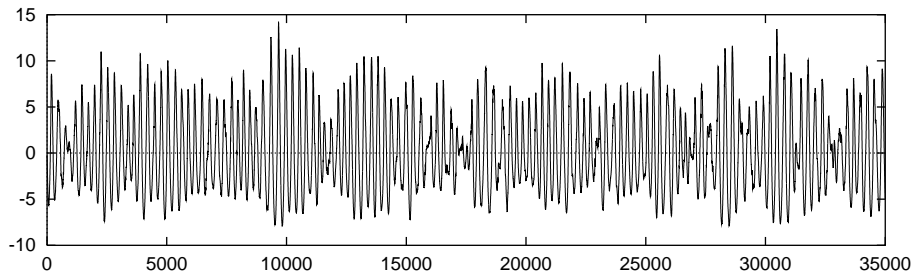}
\caption{Part of the signal analysed in this paper}
{The height
$h(t)$ in mm as a function of time $t$ in ms,
the fetch $d_f$ is 4.35 m and the wind speed
$u_w$ is 48~ms$^{-1}$.}
\label{figu48f435}
\end{figure}

The generation and dissipation of waves on the surface of the ocean under
the influence of wind is a complicated process which has been investigated
intensively over many years. With the advent of remote sensing techniques
based on microwave backscatter from the ocean surface, particular interest
has surrounded the microscale of those surface waves with wavelengths less
than approximately 0.5~m.
In the meantime wavelet methods have been developed and are now applied
to a wide range of problems see e.g. [\cite{Silverman-Vassilicos00}].
Recent years have seen the development of Continuous Wavelet Transform (CWT), filter and threshold techniques, (which corresponds to the topic of this paper) details can be found in the review paper by \cite{Farge-et-al-2011}.
There are also other applications of wavelets as reconstruction tools for synthetic turbulence methods (STM) (see for example the work of \cite{Zhou-et-al-2014}).
The study of wavelet energy spectrum for two-dimensional turbulence can also be found in [\cite{Schneider-et-al-2004}].
\\[2ex]
In this contribution we report investigations of the structure of
wind-forced microscale waves as elucidated by wavelet techniques. In
particular, we focus on the behaviour of small scale capillary waves (with
wavelengths less than approximately 20~mm) in relation to the larger scale
gravity forms.
Data used in this paper are tank waves measured
by \cite{Banner-Peirson98}.
A small part of this $2^{18}$ point long set of data is shown in
figure~\ref{figu48f435} where the displacement of the wave $h(t)$ is the quantity
measured as a function of time $t$, the fetch $d_f$ is 4.35~m and the wind speed
$u_w$ is 48~m~s$^{-1}$.
In \S \ref{sec1} we introduce general definitions and results about
wavelets. In \S\ref{sec3} we compare wavelet and Fourier spectra of the tank waves and
detail the respective contributions to the spectrum
of the signal's peaks and troughs.
In \S\ref{sec4} we define conditional spectra which we use to achieve
a better understanding
of capillary waves.
Finally in \S\ref{sec7} we compare our results to a fractal distribution
of $\Lambda$-crest.

\section{Mathematical background and definitions}
\label{sec1}

\subsection{The Wavelet transform}
\label{twofreq}

A wavelet transform of the function $h(t)$ is defined as follows
\begin{equation}
{\tilde h}(t,\tau) = {\tau}^{-1}\int h(t')
\psi^* (\frac{t'-t}{\tau})\, dt',
\end{equation}
where $\psi(t)$ is the mother wavelet, $^*$ indicates conjugate value,
${\tilde h}$ is a function of two variables $t$ and $\tau$,
$t$ is the position in the physical space (here time),
$\tau$ the wavelet scale (here it is a scale of time that is a time
lag or period).

\begin{figure}[htb]
\includegraphics[width=11.cm]{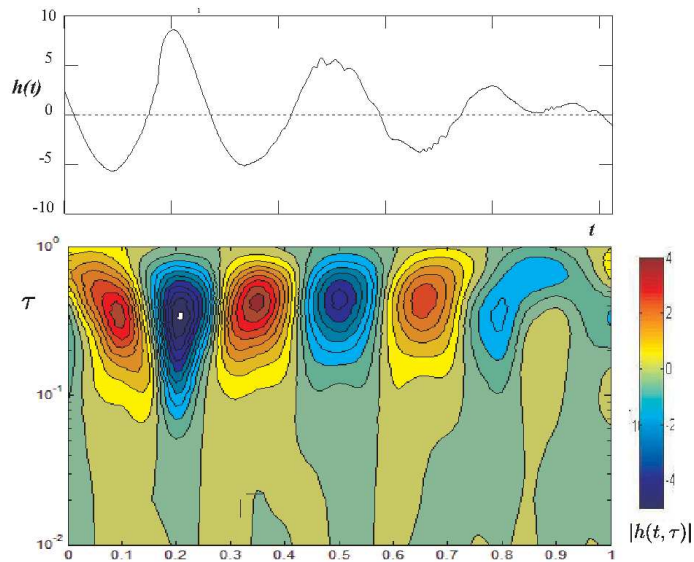}
\caption{Wavelet analysis of the 1024 first points of the wave signal
in figure~\ref{figu48f435}}
{
Upper plot is the displacement
$h(t)$ in mm as a function of time $t$ in ms,
lower plot is its wavelet transform (same as figure~\ref{fig1bis}
but viewed from a different angle).
In this latter, the scale $\tau$ is the ordinate and the time $t$ the
abscissa, curves represent iso-values of the wavelet transform modulus
$|\tilde{h}(t,\tau)|$. The mother-wavelet is the Mexican-hat.}
\label{fig1}
\end{figure}

\begin{figure}[htb]
\includegraphics[width=12.cm]{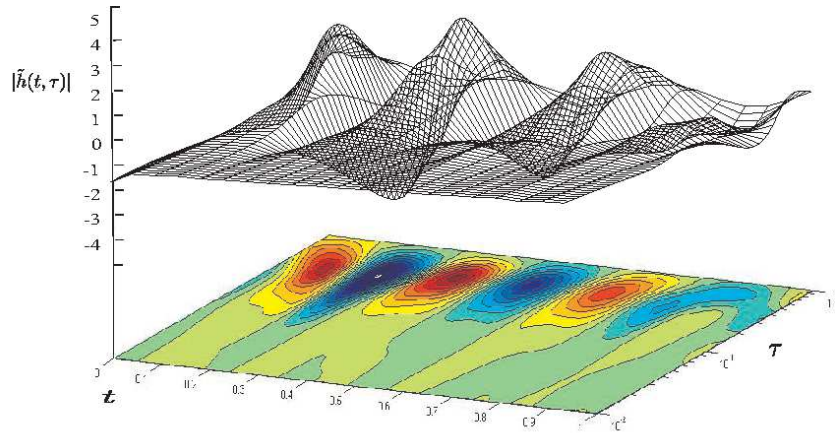}
\caption{Wavelet transform of signal in figure~\ref{fig1}}
{
Upper plot: Mexican-hat wavelet transform,
lower plot: projection of the iso-value curves.}
\label{fig1bis}
\end{figure}

The wavelet transform of $h(t)$ can be expressed as a function of its
Fourier transform
\footnote{with the standard definition $i^2= -1$}
$
\hat{h}(\omega) = \int h(t) e^{-i \omega.t} \, dt
$
and the Fourier transform
$
\hat{\psi}(\omega) = \int \psi(t) e^{-i \omega.t} \, dt
$
of $\psi$ as follows:
\begin{equation}
{\tilde h}(t,\tau)= {\tau}^{-1} \int \hat{h}(\omega) \hat{\psi}(\tau \omega)
                           e^{i \omega.\tau} \, d \omega .
\label{wavfou}
\end{equation}
$\omega$ is the frequency that is the Fourier space variable corresponding to
the time $t$ in the real space.
Figure~\ref{fig1} shows a smaller sample of the signal studied in this paper
(upper frame), the ordinate is the height of the water level ($h(t)$) and it
is measured as a function of time which we refer to as the physical space
($t$) throughout this paper.
%
%
The lower frame shows its wavelet transform, the ordinate axis holds the wavelet
scale $\tau$ and abscissa axis $t$ the physical parameter.
As shown in figure~\ref{fig1bis}, the wavelet transform
${\tilde h}(t,\tau)$ should be drawn along a third axis (upper plot),
but in order to avoid
complex 3-dimensional plots we opt for the drawing of ${\tilde h}(t,\tau)$
iso-value curves in the $(t,\tau)$ plane, that is curves defined as
$|{\tilde h}(t,\tau)| = cst$.
In practice, in this paper two mother-wavelets are used: the
Mexican-hat wavelet (see appendix~\ref{apmex}) defined as
\begin{equation}
\psi(t) = \frac{d^2}{d t^2} e^{-\frac{1}{2} t^2}
\end{equation}
and the Morlet wavelet (see \ref{apmor})
defined as
\begin{equation}
\psi(t) = e^{-\frac{1}{2}t^2} e^{it}.
\end{equation}
Both wavelets are based on the $e^{-\frac{1}{2} t^2}$ shape, the Morlet-wavelet introducing
a phase in the complex space.
This Gaussian-type shape is quite close to the actual shape of an isolated tank wave.
This is clear when comparing Figures~\ref{wiso} and \ref{phase1} to the Mexican-hat mother-wavelet given in Figure~\ref{fig18}, so we use this latter in particular to emphasize capillary and indentation effects in sections~\ref{capello} and \ref{indentations}.
The merit of each wavelet is discussed later on and in the appendices.

\subsection{Filtering}

\begin{figure}
\includegraphics[width=8cm]{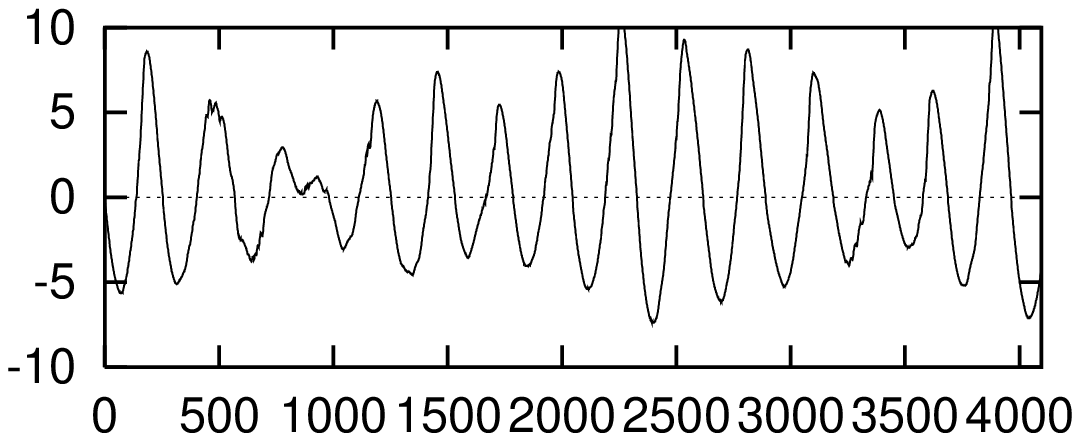}
\mbox{} \\
\includegraphics[width=8cm]{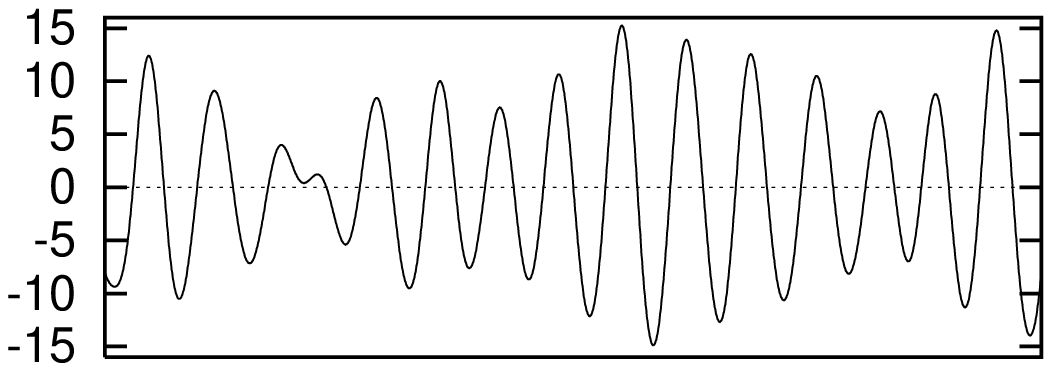}
\caption{Example of filtered signal}
{
Top,
the 4096 first points of the signal in figure~\ref{figu48f435},
bottom the scale-filtering for a scale $T_B=0.4$.}
\label{filex}
\end{figure}

Wavelet transforms enable scale-filtering of signals.
It is possible to inverse the wavelet transformation and reconstruct the
signal.
This inverse wavelet transformation is possible only when the mother
wavelet verifies some properties of integrability,
in particular that $\int \psi(t) dt =0$ (see \cite{farge92});
wavelets used in this
paper verify the required properties).
The original signal can be expressed as a linear combination of its
wavelet transform coefficients; the expression for this inversion is
\begin{equation}
h(t) = \int \int \tilde{h}(t',\tau) \, \frac{dt' \, d \tau}{\tau^2}.
\label{reconwav}
\end{equation}
Discarding certain scales in this reconstruction process defines a
scale-filter.
Such a filtering is particularly interesting for our data
as it clearly appears that they contain at least two different ranges of scales
(see section 3 and beyond):

i) the main gravity wave scale,

ii) the capillary waves and small scale indentations.
\\
As we will see in \S\ref{sec4} the
wavelet-scales involved in the main gravity waves are
clearly an order of magnitude or two larger than those involved in the
capillarity effects and small-scales indentations.
We propose to define the small-scales filtered signal $h_{T_B} (t)$ as follows:
\begin{equation}
h_{T_B}(t) = \tilde{h}(t, T_B)
\end{equation}
This is equivalent to using the filter $\tau^2 \delta (\tau-T_B)\delta(t'-t)$
in the integration of the right hand side of (\ref{reconwav}).
Figure~\ref{filex} shows a portion of the signal $h(t)$ in
figure~\ref{figu48f435} and $h_{T_B}(t)$ the result of the filtering out of
scales smaller and larger than $T_B$.


\subsection{Local spectrum}

From the wavelet transform of $h(t)$ we can compute the local energy density
[\cite{farge92}]
\begin{equation}
\tilde{E}(t,\tau) = |\tilde{h}(t,\tau)|^2 \, \tau^2.
\label{loc1}
\end{equation}
Equation (\ref{loc1}) gives information about energy density
associated with scale $\tau$ and localised at time
$t$. The global wavelet spectrum $\tilde{E}(\tau)$ is the sum of all
these local wavelet spectra,
\begin{equation}
\tilde{E}(\tau) = \int \tilde{E}(t,\tau) \, dt.
\label{glo}
\end{equation}
It relates to the Fourier spectrum:
\[
\tilde{E}(\tau) = \int |\int \hat{h}(\omega) \hat{\psi}(\tau \omega)
                     e^{i \omega t} \, d \omega |^2 \,dt.
\]
With superscript $^*$ indicating conjugate value,
\[
\tilde{E}(\tau) = \int \int \int \hat{h}(\omega) \hat{\psi}(\tau \omega)
                              \hat{h}^*(\omega')\hat{\psi}^*(\tau \omega')
                    e^{i(\omega-\omega')t} \, d\omega \, d\omega' \,dt,
\]
that is:
\[
\tilde{E}(\tau) = \int \int \left \{ \int e^{i(\omega-\omega')t} dt \right \}
                         \, \hat{h}(\omega) \hat{\psi}(\tau \omega)
                         \hat{h}^*(\omega')\hat{\psi}^*(\tau \omega')
                         \, d \omega \, d \omega',
\]

\[
\tilde{E}(\tau) = \int \int \delta(\omega-\omega')
                         \hat{h}(\omega) \hat{h}^*(\omega')
                         \hat{\psi}(\tau \omega) \hat{\psi}^*(\tau \omega')
                         \, d \omega \, d \omega'
\]
and eventually
\begin{equation}
\tilde{E}(\tau) = \int |\hat{h}(\omega)|^2 |\hat{\psi}(\tau \omega)|^2
\, d \omega.
\label{eqspew}
\end{equation}
The wavelet spectrum therefore appears as an average of Fourier spectra
$E(\omega)=|\hat{h}(\omega)|^2$ weighted with the wavelet
term $|\hat{\psi}(\tau \omega)|^2$.
\\[2ex]
If the Fourier spectrum has a power law over a range of frequencies,
i.e. $E(\omega) \sim \omega^{-p}$ when $\omega \to \infty$,
the change of variable $k'= \tau \omega$ can be used to obtain the power
law of $\tilde{E}(\tau)$ when $\tau \to 0$. Indeed with this change of
variables,
\begin{equation}
\tilde{E}(\tau) = 2 \tau^{-1} \int E(\frac{k'}{\tau})
               |\hat{\psi}(k')|^2 \, d k',
\label{chgvar}
\end{equation}
and in the limit $\tau \to 0$ we have
\[
\tilde{E}(\tau) = 2 \tau^{p-1} \int {k'}^{-p}
               |\hat{\psi}(k')|^2 \, d k',
\]
that is:
\begin{equation}
\tilde{E}(\tau) \sim \tau^{p-1}.
\label{power}
\end{equation}
This result is independent of the choice of the wavelet $\psi$
but due to the fact that the integral in (\ref{chgvar}) is in practice
taken over a finite range of $k'$, some wavelets give better results than
others. We use here the Morlet wavelet to educe cut-off scales
(see appendix~\ref{apmor}) and
the Mexican-hat wavelet to educe power law spectra (see appendix~\ref{apmex}).

\subsection{Conditional spectrum}
\label{secd4}

The definition of the wavelet spectrum allows definitions of
conditional spectra.
A conditional spectrum is defined as the integration of the wavelet local
spectrum over a given region $(V_c)$ in the physical space where the required condition
is met.
\begin{equation}
\tilde{E}_c(\tau) = \frac{1}{V_c} \int_{V_c} \tilde{E}(t,\tau) \, dt.
\end{equation}
In our case of analysing tank waves, we can define different conditions
to educe the contribution of the peaks, troughs and different parts of the
elementary waves to the global spectrum.
The variation of $\tilde{E}_c(\tau)$ according to the definition of $V_c$
gives information on how the global spectrum relates to the different
regions in the physical space.
For instance contribution to the spectrum of peaks in the data
can be estimated by setting the condition
\begin{equation}
h(t) > \epsilon,
\label{cond1}
\end{equation}
as in \S\ref{sec3}-\ref{peaks}.
The higher the value of $\epsilon$
the more affected is the conditional spectrum by the peaks.
Similarly, the condition
\begin{equation}
h(t) < \epsilon
\label{cond2}
\end{equation}
gives information on how
the global spectrum relates to the trough regions as in \S\ref{sec3}-\ref{troughs}.
It is also possible to condition the spectrum on different
parts of each elementary wave as shown in \S\ref{sec4}.

\section{Spectrum of the tank waves}
\label{sec3}

\subsection{Wavelet spectrum}

\begin{figure}
\hspace*{2.5cm}
\includegraphics[height=5.4cm]{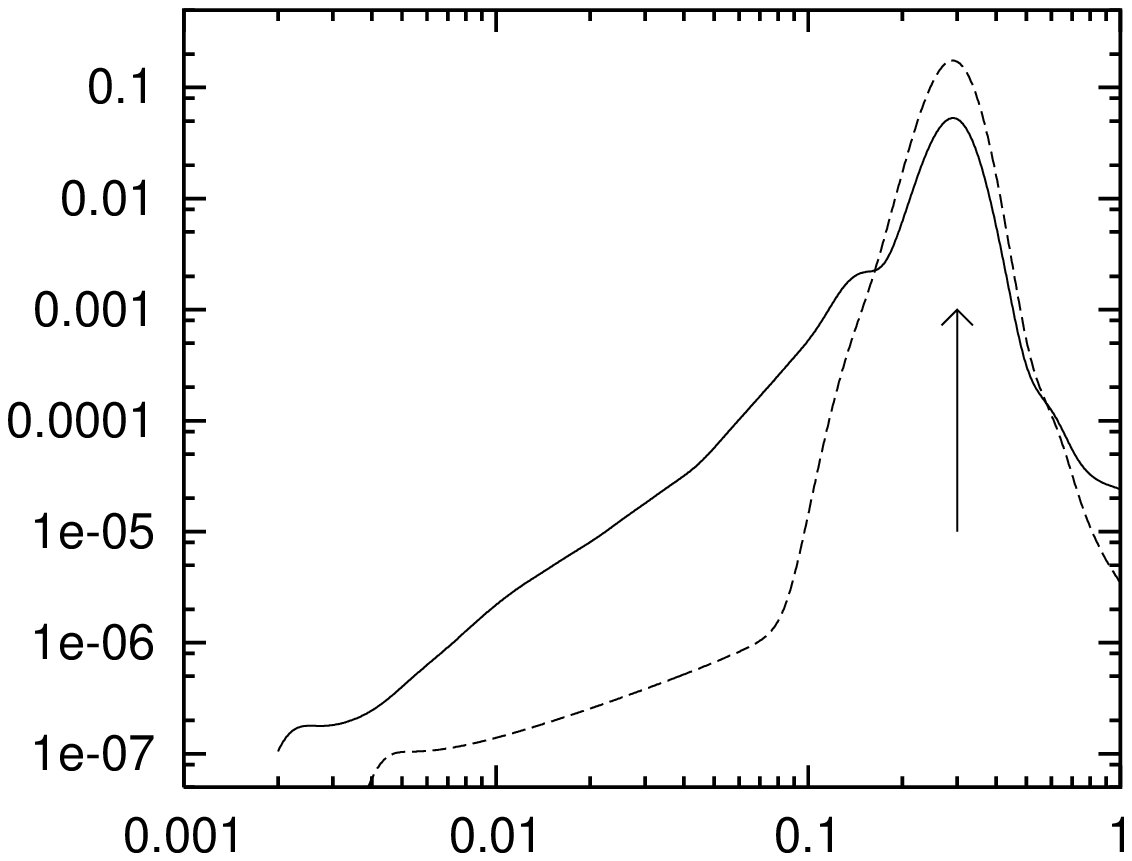}
\caption{Wavelet spectrum of a sample of
the signal shown in figure~\ref{figu48f435}}
{
$\tilde{E}(\tau)$ using the Morlet wavelet for a
131072 point long data sample;
solid line: entire signal, dash line: filtered signal $h_{T_B} (t)$.}
\label{figspecwf}
\end{figure}
Figure~\ref{figspecwf} shows the wavelet spectrum $\tilde{E}(\tau)$ of the
displacement $h(t)$ as a function of the scale $\tau$.
Solid lines correspond to the entire 131072 point long data sample of
the signal shown in figure~\ref{figu48f435} and dash lines to the same
signal $h_{T_B} (\tau)$ filtered at the scale $T_B=0.4$.
Note how the Morlet wavelet spectrum can be used to
educe characteristic frequencies, which can of course also be done by Fourier methods.
The filtered signal
has virtually no wavelet intensity at small scales.
All the wavelet intensity is focused on the scale 0.3
corresponding to the distance $2 \lambda$
between two minima or two maxima of the
signal. Direct measurement of the average value of $\lambda$ over the entire signal
gives $\lambda \approx 0.13$.
Filtering the signal does not alter the position of its zero-crossings but it
drastically changes the shapes of the waves
mainly by smoothing them and naking them more top-bottom symmetric.
(see figure~\ref{filex}).
Hence one can conclude from figure~\ref{figspecwf}
that the small-$\tau$ part of the wavelet spectrum ($\tau < 0.13$)
is mainly due to the shape of each individual wave,
a fact confirmed by the analysis of an isolated wave in \S\ref{sec4}.

\begin{figure}
\includegraphics[height=4.5cm]{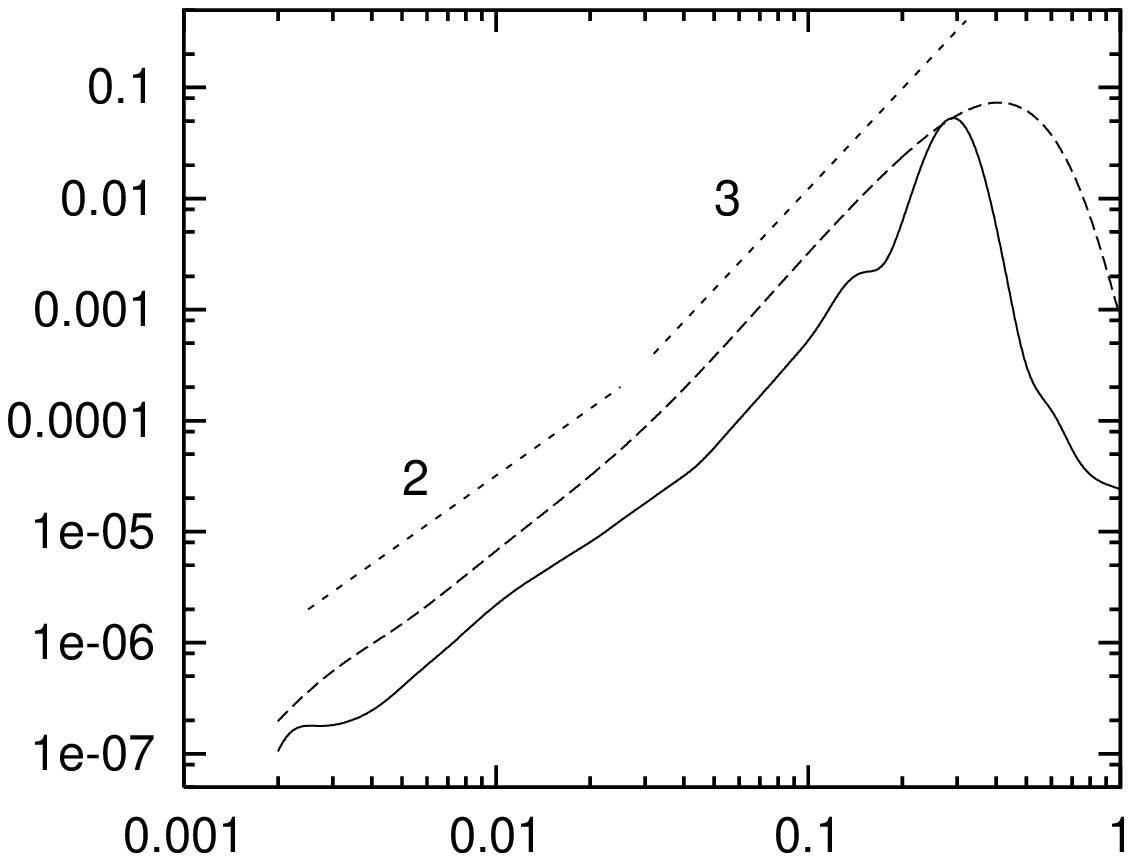}
\includegraphics[height=4.5cm]{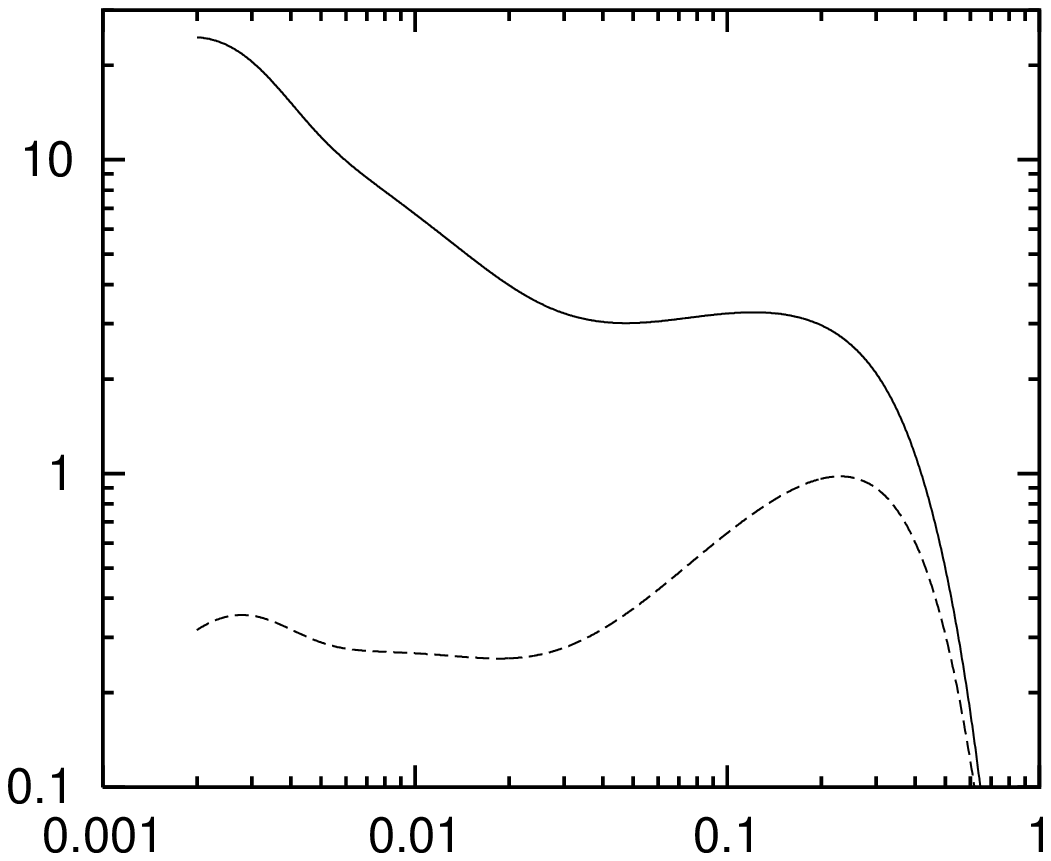}
\\
\hspace*{2cm} a) \hspace*{6.5cm} b)
\\[2ex]
\includegraphics[height=4.5cm]{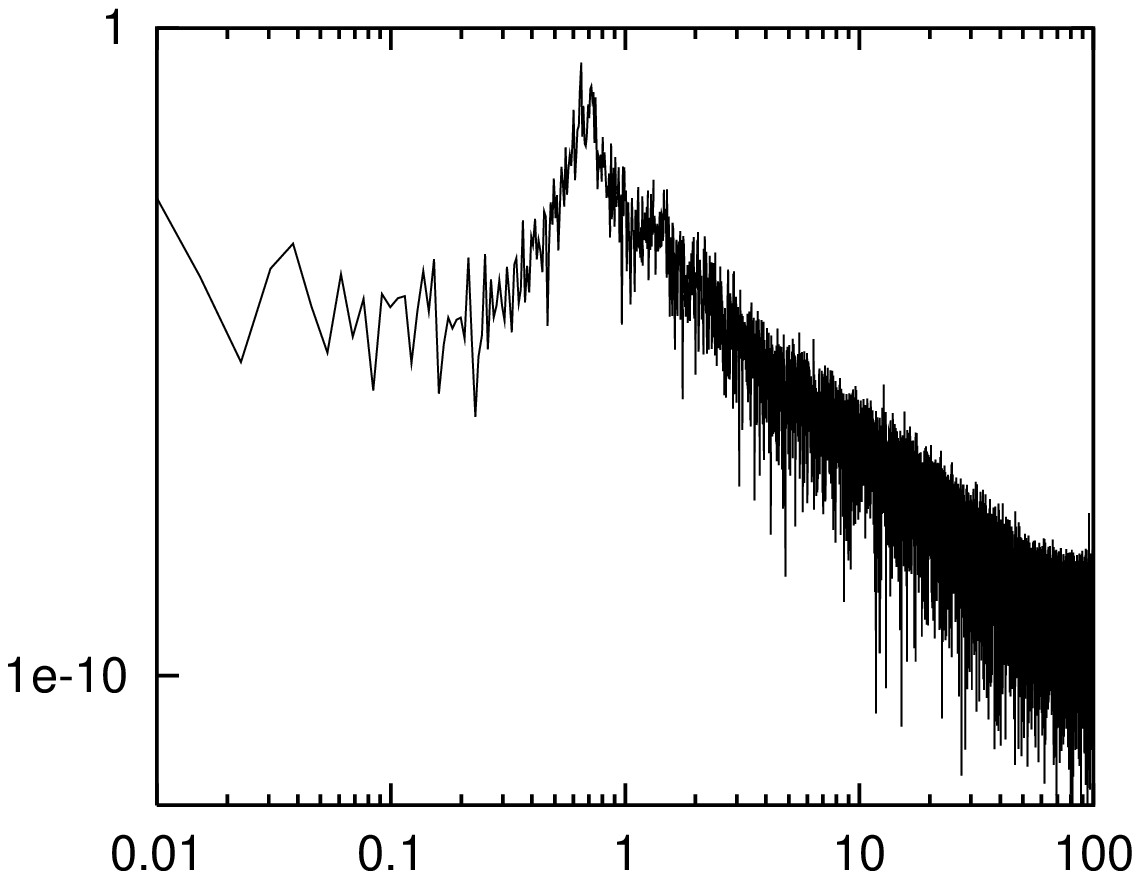}
\includegraphics[height=4.5cm]{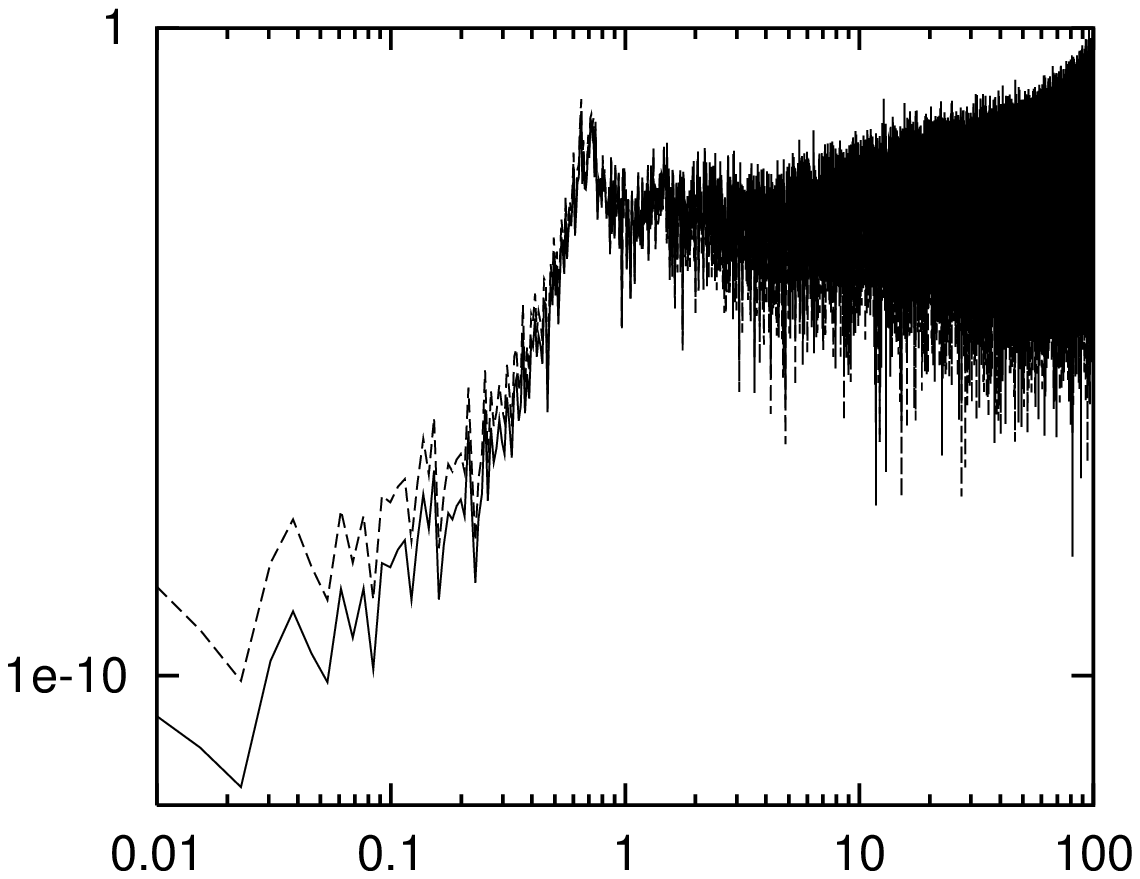}
\\
\hspace*{2cm} c) \hspace*{6.5cm} d)
\caption{Wavelet and Fourier spectra of a 131072 point long data sample of
the signal shown in figure~\ref{figu48f435}}
{
a) Wavelet spectra $\tilde{E}(\tau)$
as a function of $\tau$ for Mexican-hat (upper
curve) and Morlet wavelet (lower curve).
b) $\tau^3 \tilde{E}(\tau)$ (upper curve) and $\tau^{2.3}
\tilde{E}(\tau)$ (lower curve) as functions of $\tau$ for the spectrum
based on the Mexican-hat wavelet.
c) Fourier spectrum $E(\omega)$ as a function of $\omega$ for the
same data sample.
d) $\omega^4 E(\omega)$ as a function of $\omega$ (upper curve)
and $\omega^3E(\omega)$ (lower curve)
- $\omega \approx \frac{2 \pi}{\tau}$.}
\label{figspecw}
\end{figure}

Figures~\ref{figspecw}a,b show the wavelet spectrum $\tilde{E}(\tau)$ of the
signal in figure~\ref{figu48f435} based on the Mexican-hat wavelet which is
suitable
for educing power laws.
It is not possible to conclude on the existence of power laws for these
data, the range of scales being very small, but it is possible to educe
two different regions.
Asymptotically for $\tau$ small we seem to have
$\tilde{E}(\tau) \sim \tau^{2.3}$ (see figure~\ref{figspecw}b)
in the range $\tau_c<\tau<\tau_{trans}$ where $\tau_c =0.002$
is the lower cut-off
scale of the data\footnote{
Actually it is 0.001, but the smallest time-scale which can be observed by the
wavelet cannot be smaller than twice the lower cut-off scale.}
and $\tau_{trans}$ the upper limit for this first region
(here $\tau_{trans} \sim 0.03$).
For larger scales ($\tau > \tau_{trans}$) we seem to have $\tilde{E}(\tau) \sim
\tau^{3}$ (see figure~\ref{figspecw}b),
which according
to (\ref{power}) corresponds to a Fourier spectrum $E(\omega) \sim
\omega^{-4}$, a result consistent
with the existence of $\Lambda$-crests (that is crests with
discontinuity in slope) all of same duration.
In \S\ref{sec3}\ref{fourier} we show that the accuracy on $\tilde{E}(\tau)$ is
by far superior to that on $E(\omega)$.

\subsection{Comparison with Fourier spectrum analysis}
\label{fourier}

The Fourier spectrum of $h(t)$ is classically defined as
\begin{equation}
E(\omega) = |\hat{h}(\omega)|^2.
\end{equation}
The Fourier spectrum of the signal in figure~\ref{figu48f435} is given in
figure~\ref{figspecw}c. It contains a lot of noise which is not the case of
the wavelet spectra curves $\tilde{E}(\tau)$ (figure~\ref{figspecw}a).
The peak is reached at
$\omega=13.5$, that is $\tau=2 \pi / \omega = 0.465$,
and a slope may be observed in the range 15--300.
Figure~\ref{figspecw}d shows both $\omega^4 E(\omega)$ and
$\omega^3 E(\omega)$.
Due to the noise
it is difficult to decide whether $E(\omega) \sim \omega^{-4}$ or
$E(\omega) \sim \omega^{-3}$.
It seems however that $E(\omega) \sim \omega^{-4}$ in the range 15--300 and
$E(\omega) \sim \omega^{-3}$ in the range 300--1000.
We show in section 4 that
this value of $\omega \sim 300$ or
$\tau \sim 0.003$ is rather close to the scale ($\sim 0.010$) of
very small perturbations observed on certain parts of the signal.

\subsection{Energy spectrum of peaks in the signal}
\label{peaks}

%
\begin{figure}
\includegraphics[height=4.5cm]{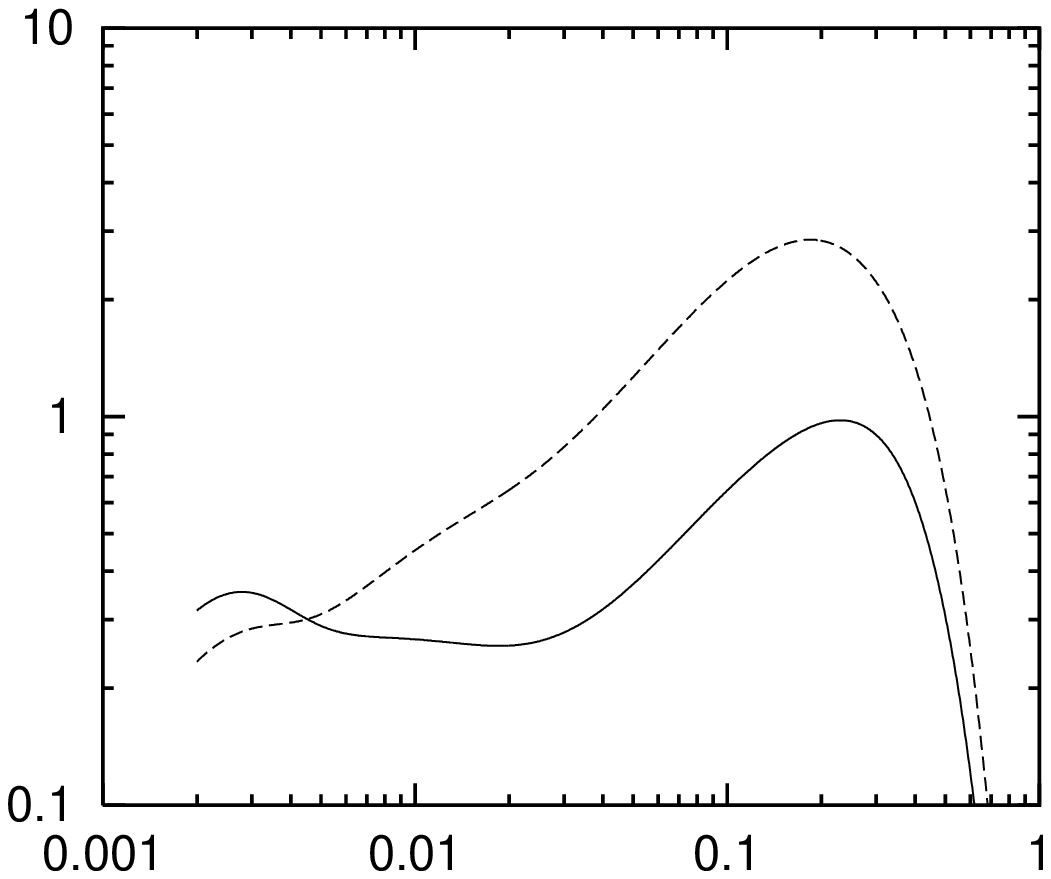}
\includegraphics[height=4.5cm]{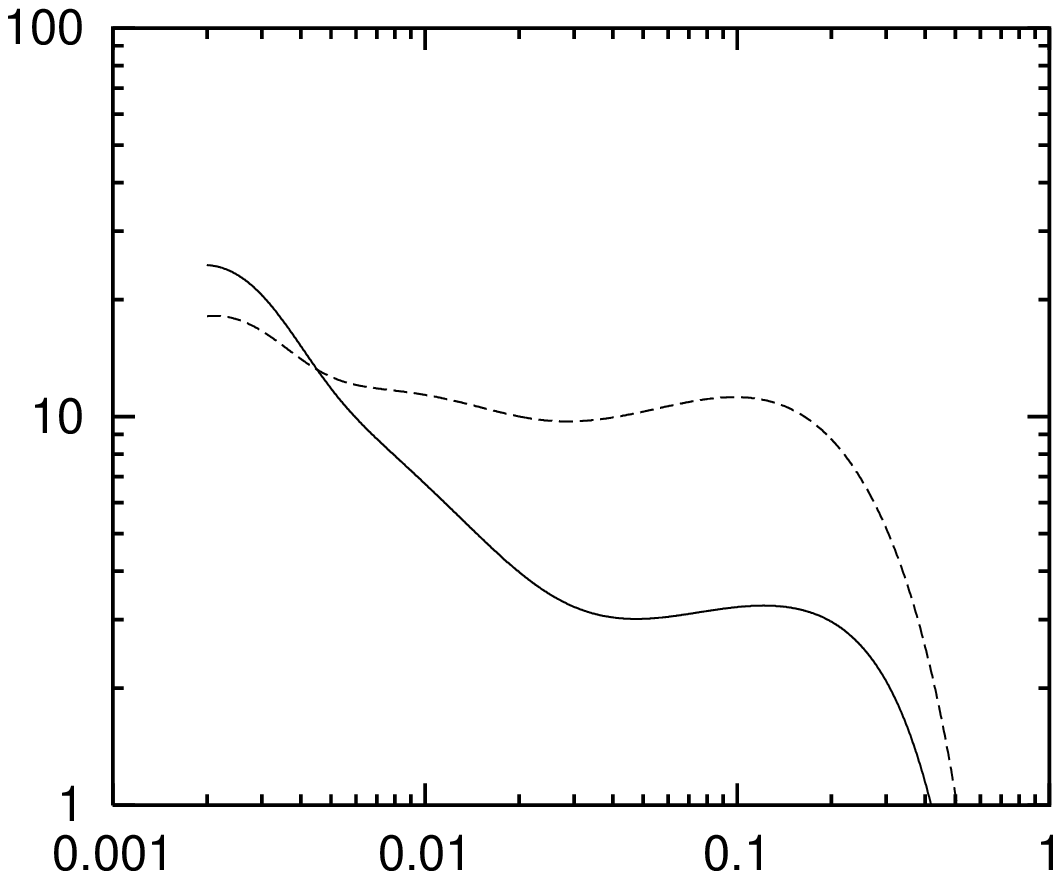}
\\
\hspace*{ 2cm} a)
\hspace*{ 6.5cm} b)
\caption{Conditional spectra based on the condition $h(t) > \epsilon$}
{
From the signal in figure~\ref{figu48f435},
power spectra are based on 131072 points and we use the Mexican-hat wavelet.
a) $\tau^{-2.3} \tilde{E}_c(\tau)$ against $\tau$
for the entire signal (lower curve) and for $\epsilon = 5$,
b) $\tau^{-3} \tilde{E}_c(\tau)$ against $\tau$
for the entire signal (lower curve) and for $\epsilon = 5$.}
\label{figspecum}
\end{figure}
%
%
In this section, condition (\ref{cond1}) is used to compute conditional
spectra for the 131072 first points of the signal in figure~\ref{figu48f435}.
Figure~\ref{figspecum}a shows the compensated non conditioned spectrum
$\tau^{-2.3} \tilde{E}(\tau)$ (lower curve) and
the compensated conditional spectrum (upper curve) based on the criterion
$h(t) > 5$.
This latter spectrum is associated with peaks: from
figure~\ref{figu48f435}, one can see that for $\epsilon \ge 5$,
the conditioned signal is just a sum of isolated wave-crests.
It is clear from figure~\ref{figspecum}a that peaks play no part in the power
law $\tilde{E}(\tau) \sim \tau^{2.3}$ observed for the small scales
(i.e. $\tau <\tau_{trans}$) in the non conditioned signal.
Figure~\ref{figspecum}b shows the compensated spectrum
$\tau^{-3} \tilde{E}(\tau)$
obtained from the entire signal and from the condition $h(t) > 5$.
The conditional spectrum is closer to $\tilde{E}(\tau) \sim \tau^{3}$ down to
small scales $\tau < \tau_{trans}$. A transition still appears at the scale
$\tau = \tau_{trans}$ but it is much less stressed than in the case of
the entire signal.

From this spectral analysis we can conclude that the signal's peaks are close to
discontinuities in slope. Indeed, it is known from Fourier
analysis that such discontinuities are characterised by a $\omega^{-4}$ spectrum.
This indicates that for $\tau > \tau_{trans}$, the signal's spectrum
is dominated by wave-crests.
(See \S\ref{sec4}\ref{isocrest} for the analysis of an isolated crest.)

\subsection{Energy spectrum of the troughs in the signal}
\label{troughs}

\begin{figure}
\includegraphics[height=4.5cm]{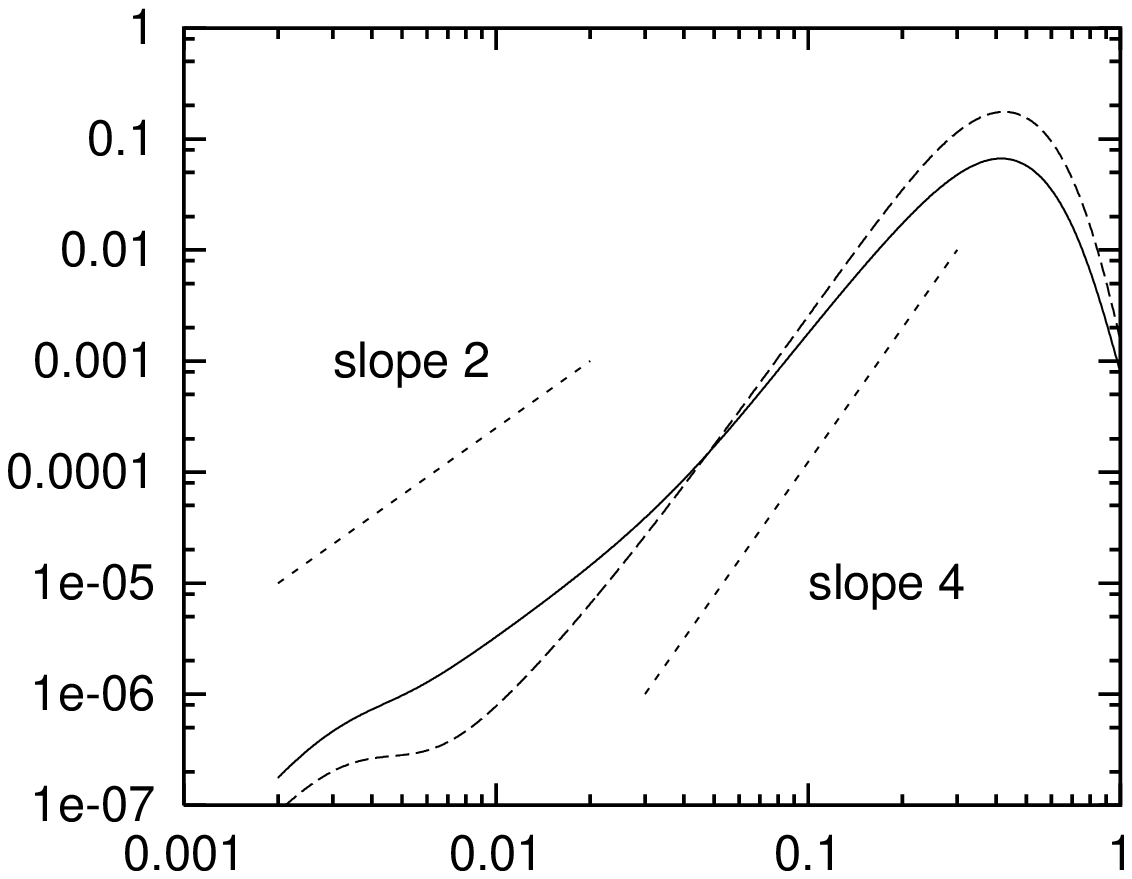}
\includegraphics[height=4.5cm]{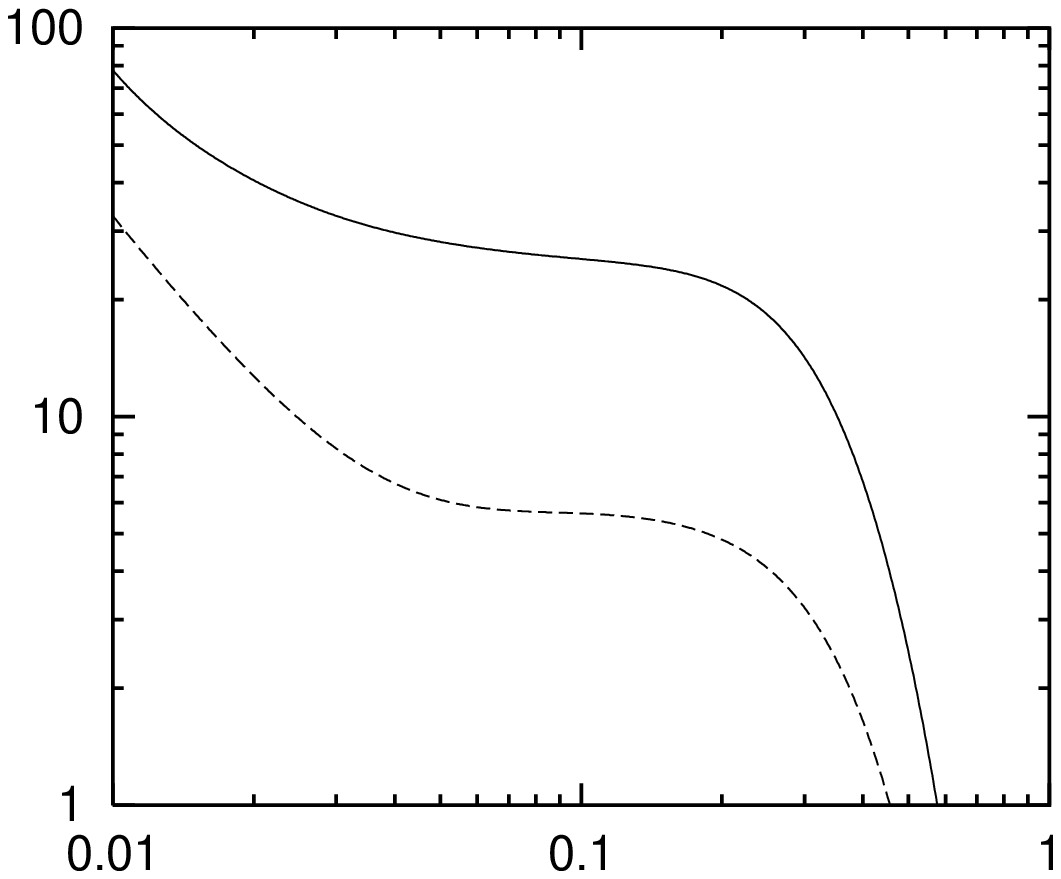}
\\
\hspace*{2cm} a) \hspace*{6.5cm} b)
\caption{
$\tilde{E}(\tau)$, spectra conditioned on the troughs: $h(t) < \epsilon$}
{
For
$\epsilon = 0$ solid line and $\epsilon = -5$ dash line, a) wavelet spectrum,
b) $\tau^{-3.5}\tilde{E}(\tau)$ for the spectrum conditioned on $h(t) <0$
(solid line) and $\tau^{-4}\tilde{E}(\tau)$ for the spectrum conditioned on
$h(t) < -5$ (dash line), spectra are measured on a $2^{17}$ point long
sample and based on the Mexican-hat wavelet.}
\label{figspetrou}
\end{figure}

Using condition (\ref{cond2}) we now analyse the local spectrum of
the troughs.
The conditional wavelet spectra shown in
figure~\ref{figspetrou} give some idea of the shape of the local spectrum associated
with troughs.
Again there are two ranges of scales $\tau$ separated by a value
$\tau_{trans}$, and the lower $\epsilon$ - i.e. the closer to the troughs - the
smaller $\tau_{trans}$. Figure~\ref{figspetrou}b shows the compensated spectra
$\tau^{\alpha} \tilde{E}(\tau)$.
It seems that troughs do not have a $\tau^3$ wavelet spectrum
but are closer to a $\tau^4$ power law in the range $\tau_{trans} <\tau< 0.2$.
Trough and peak regions have different contributions to the global
energy spectrum.

\section{Capillary effects}
\label{sec4}

\subsection{Analysis of an isolated wave}
\label{isocrest}

%
\begin{figure}
~\hspace*{0.3cm}
\includegraphics[height=4.5cm]{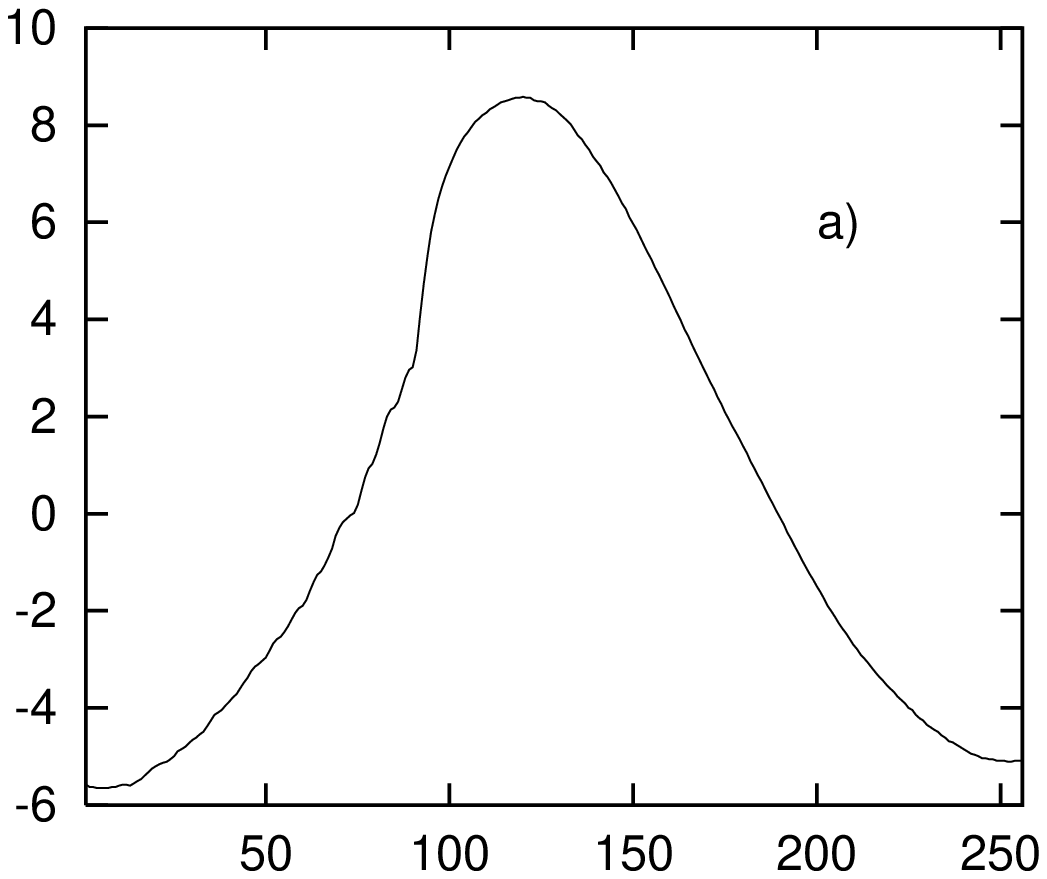}
\includegraphics[height=4.5cm]{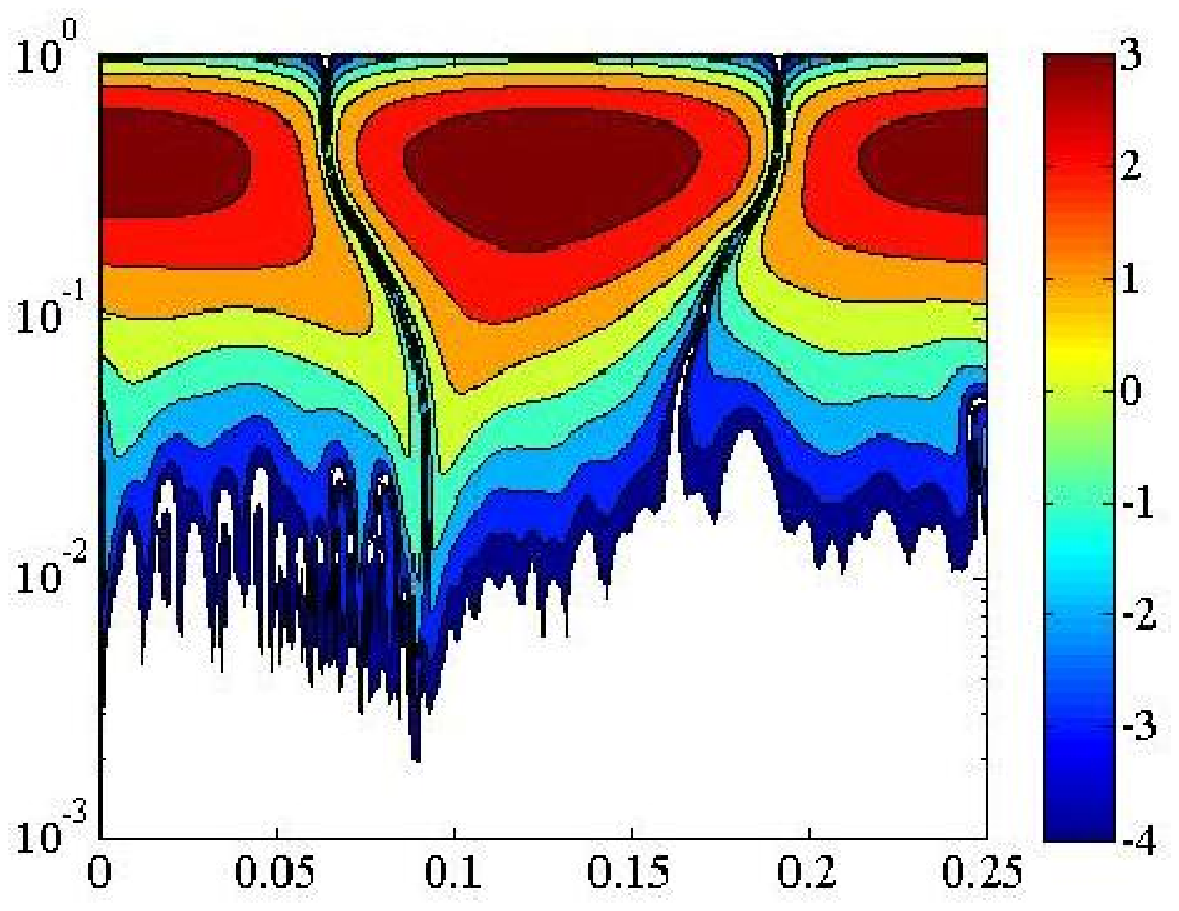}
\\
\hspace*{1.cm} a) \hspace*{6.cm} b)
\\[2ex]
\includegraphics[height=4.5cm]{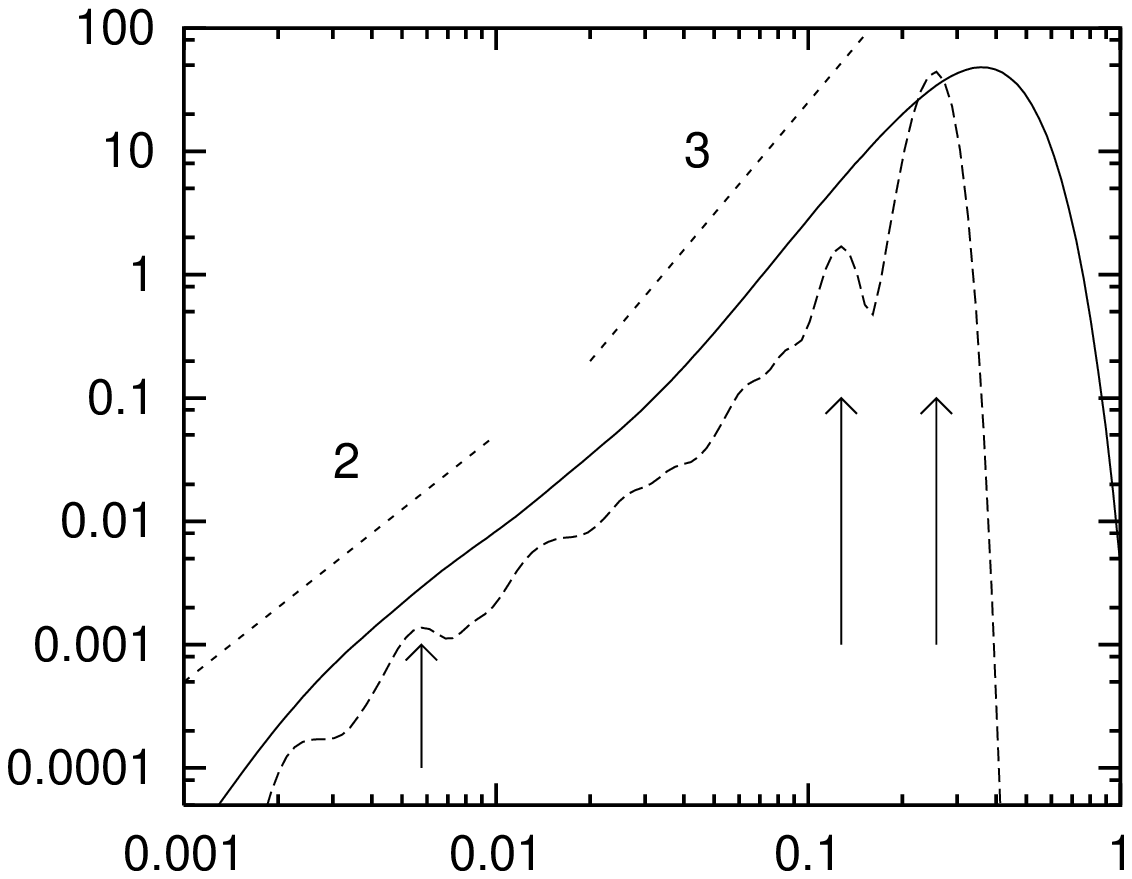}
\includegraphics[height=4.5cm]{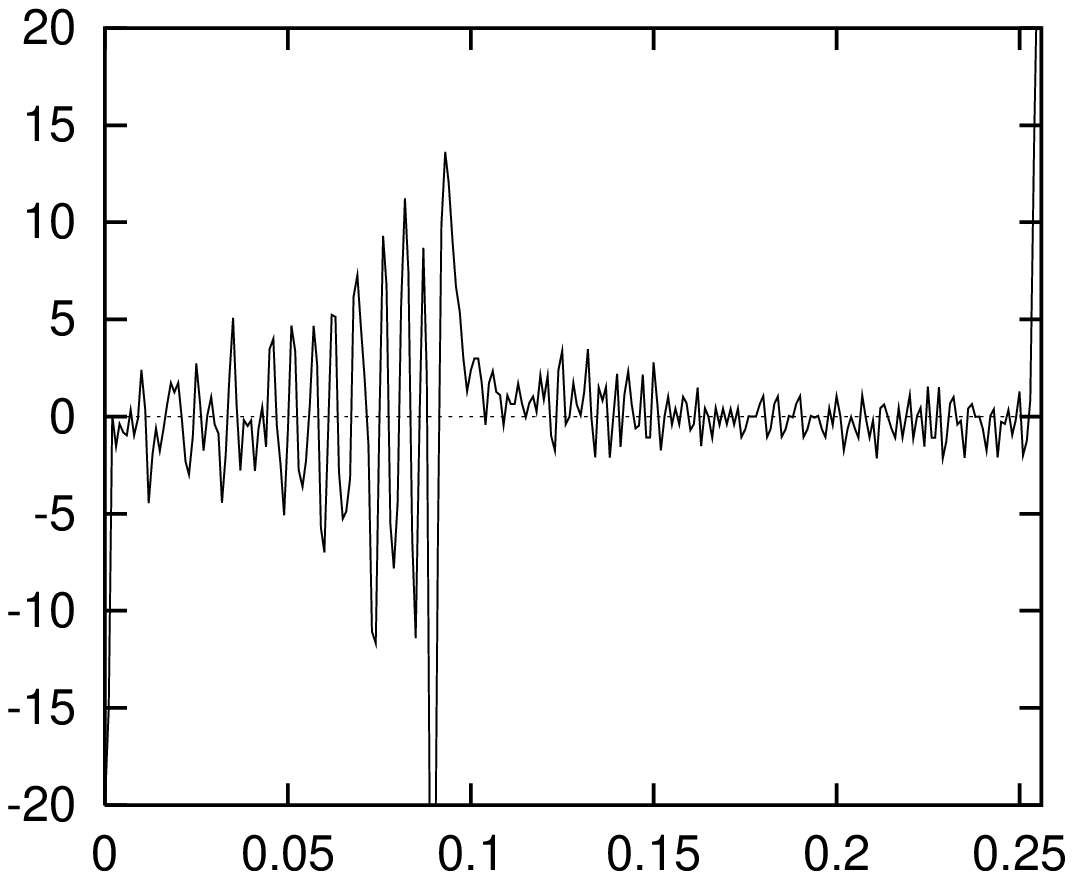}
~\\
\hspace*{1.cm} c) \hspace*{6.cm} d)
\caption{Isolated wave, 256 points}
{
a) Displacement $h(t)$ measured as a function of $t$
given in milliseconds.
b) Wavelet transform of the displacement, plot in semi-log, the
y-axis holds the scale parameter $\tau$ and the x-axis holds the physical parameter $t$
given in seconds.
Curves are iso-values of the wavelet energy $| \tilde{h}(t,\tau) |= cst$.
c) Wavelet spectra based on Mexican-hat (solid line) and
Morlet (dashed line) wavelets.
d) $\tilde{h}(t,\tau=0.005)$ using the Mexican-hat wavelet.}
\label{wiso}
\end{figure}

We can limit our study to an isolated wave such as the one in
figure~\ref{wiso}a.
A common feature of water waves is the small scale parasitic capillary waves
which are present on the forward faces.
Water waves with lengths less than about 10mm are called capillary waves and
those riding on the forward faces of the larger gravity forms are called
parasitic capillaries because they derive their energy from the larger wave.
These capillary waves are clearly educed with the use of wavelet transform
as can be seen in figure~\ref{wiso}b.
The wavelet transform clearly localises a main discontinuity at
$t=0.09\pm 0.001$.
Capillary waves appear as blobs of wavelet intensity in the interval $0<t<0.09$
(i.e. on the forward face), and
they are associated with a characteristic period (scale $\tau$)
of the order of $\tau \sim 0.01$.

Figure~\ref{wiso}c shows the wavelet spectra based on Mexican-hat
(solid line) and Morlet (dashed line) wavelets
for the signal in figure~\ref{wiso}a.
The Morlet wavelet is good at educing cut-off scales and the wave
frequency.
The wave period and half-period are shown by the arrows at
0.257 and 0.127, the third arrow points to the capillary wave period
0.0056. These values are consistent with an examination of
figure~\ref{wiso}b, i.e. the capillary wave period is
1/50 of the characteristic period of the larger scale gravity wave
which appears as large scale blobs at the top of figure~\ref{wiso}b.
%
The Mexican-hat wavelet is clearly better at educing the
spectral power law $\tilde{E}(\tau) \sim \tau^{p}$. Slopes 2 and 3 are
drawn in figure~\ref{wiso}c giving an indication of possible power
laws but ranges are too small to conclude here.

The scale associated to the capillary effect being clearly educed,
figure~\ref{wiso} shows the filtered the signal for this scale $\tau=0.005$.

\subsection{Wavelet spectra conditioned on phases}
\label{phases}

\begin{figure}
\hspace*{2cm}
\includegraphics[height=5cm]{{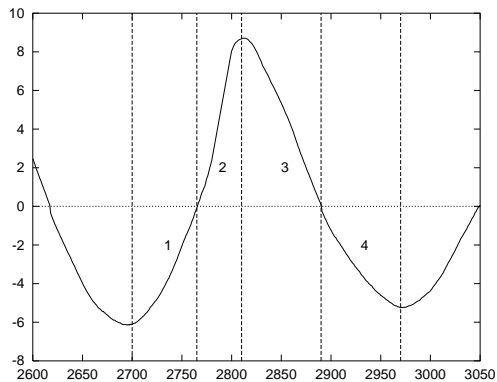}}
\caption{Definition of the elementary parts of an isolated wave
on an example of a wave different from figure~\ref{wiso}}
\label{phase1}
\end{figure}

It is also possible to condition the spectrum on the different
part of each elementary wave as shown in figure~\ref{phase1}.
Wind blows from 4 to 1; four conditional spectra can be defined
corresponding to each region 1,2,3,4 of the wave. The wave breaking
occurs in region 2 and capillary effects are observed in region 1 and 4.

An especially important feature of our data is the propagation of
capillary waves in the highly sheared current of the viscous boundary
layer.
Parasitic capillary waves are generated by small-scale gravity waves.
Extensive theoretical and numerical investigations by \cite{Longuet-H95} (see references there in)
has revealed the mechanisms linking the formation of small
scale gravity waves with the parasitic capillary waves located downwind of
their crests. The combined effects of surface tension and high curvature at
the crests of small gravity waves form a localised moving disturbance at the
surface. This disturbance generates capillary waves upstream and gravity
waves downstream due to the dispersion relation near the gravity/capillary
transition.

The breaking of small scale gravity waves results in the generation of
capillary waves.
During the breaking process, wave energy is not only dissipated by
subsurface turbulence but is also radiated from the breaking region
[\cite{rm90}].
The relationship between these generated waves and the underlying
gravity wave form is probably most clearly revealed by the laboratory
experiments of \cite{bf85}. Notably, the frequencies of these
waves are substantially higher than the gravity wave and the influence of
capillarity makes these waves highly dissipative and, therefore, short
lived. Their random formation and rapid dissipation makes monitoring their
motion exceedingly difficult.
\begin{figure}
\includegraphics[height=4.3cm]{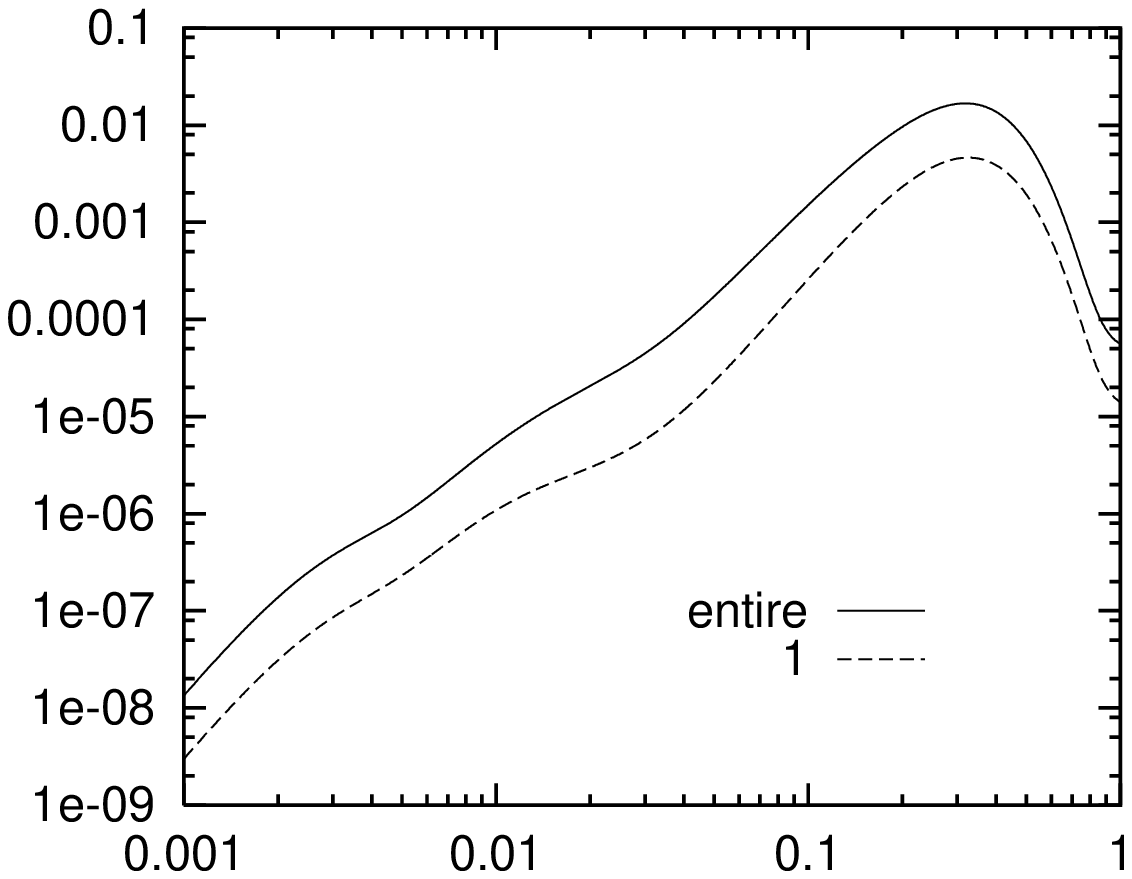}
\includegraphics[height=4.3cm]{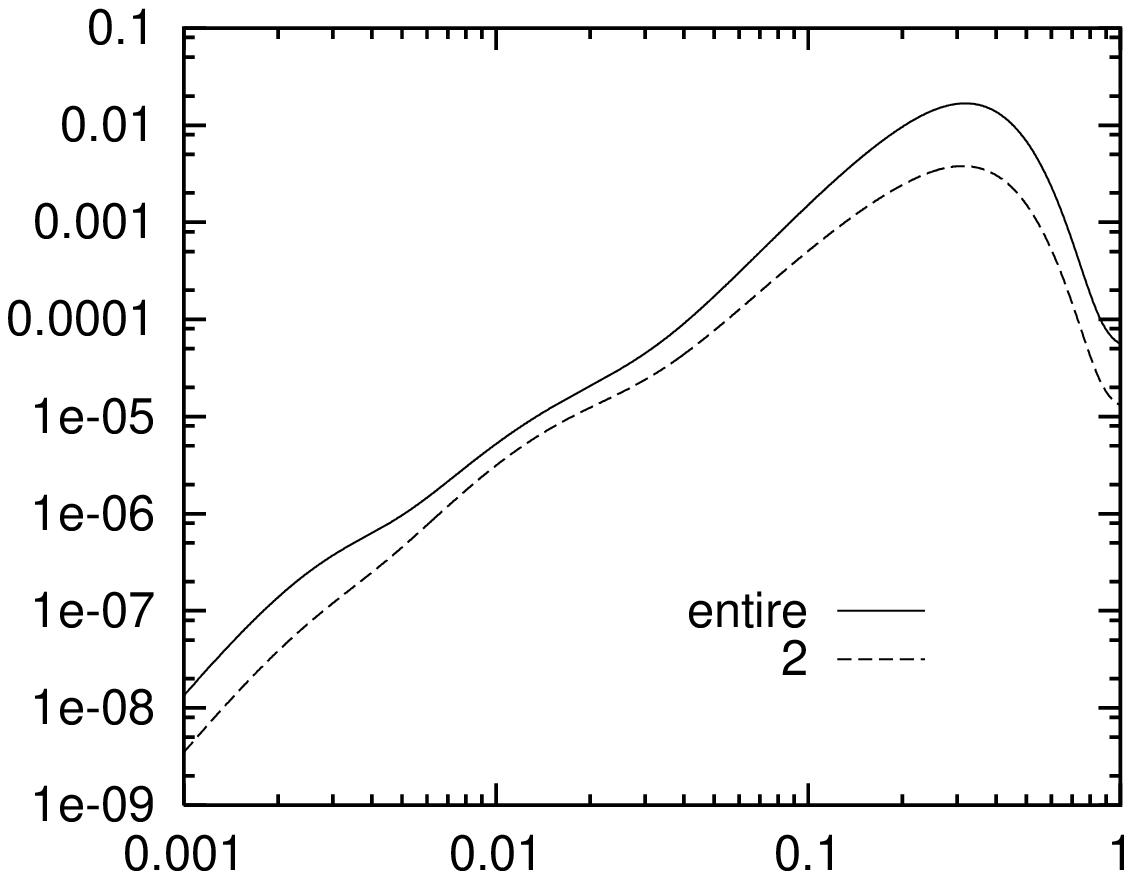}
\\
\hspace*{2 cm} a) \hspace*{6.cm} b)
\\[2ex]
\includegraphics[height=4.5cm]{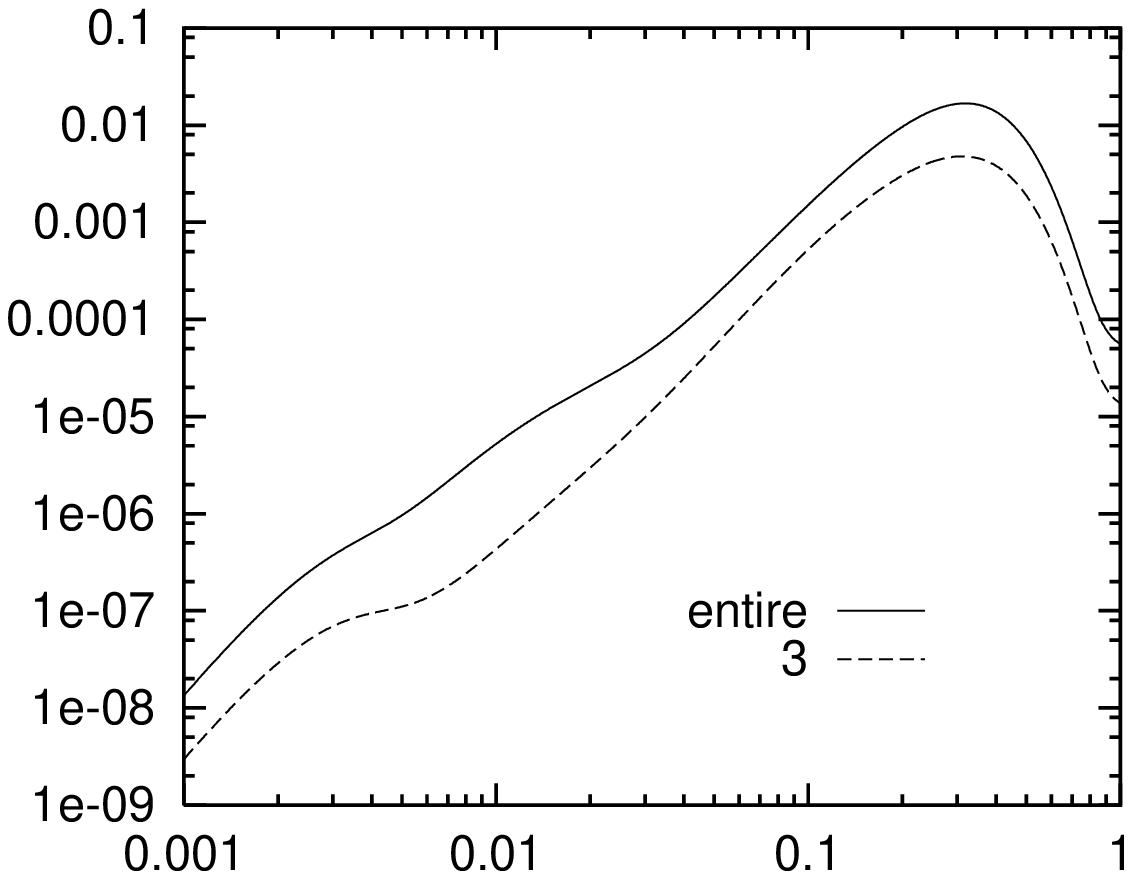}
\includegraphics[height=4.5cm]{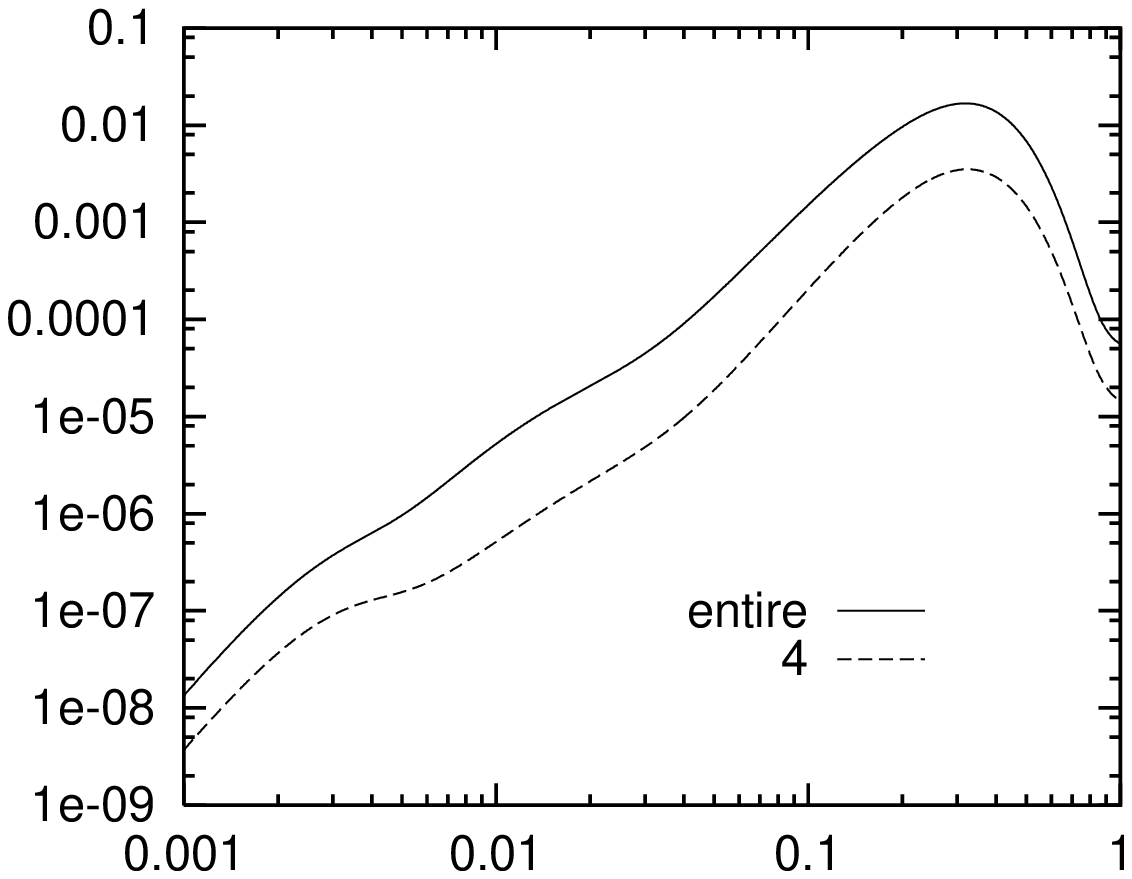}
~\\
\hspace*{2 cm} c) \hspace*{6.cm} d)
\caption{Conditional spectra, 1,2,3,4 refer to conditions described
in figure~\ref{phase1}}
\label{phafig1}
\end{figure}
\\[2ex]
Uncertainty existed as to the direction of propagation of waves generated
by breaking as it has been suggested that the wind drift at the crest was
greatly intensified by local tangential stresses. Recent work
by \cite{Banner-Peirson98} has shown that relative to the moving wave form (except in
the immediate vicinity of the spilling region) the mean wind drift is
approximately $0.3(\pm-0.1)u_*$ and transport is in an upwind direction.
Whilst the sources of capillary wave energy have been identified, very few
investigations have endeavoured to distinguish between their relative
contributions to surface wave energy. In particular, the investigations of
\cite{bf85} have ignored the role of
wind, yet the wind itself is plainly able to generate capillary ripples.
Furthermore, the conclusions of these two detailed studies is that a strong
relationship exists between the larger scale wave and the high frequency
motions which occur on its surface. Yet the spectral relationships
determined take no account of these directly.
\\[2ex]
Here,
the high resolution wave probe record is examined using
wavelet techniques with a view to determining the sources and relative
contributions of capillary wave energy along representative wind wave forms.
Figures~\ref{phafig1} show comparisons of the entire spectrum, upper curves
in each plot, with conditional spectra based on sector~1,2,3,4 defined in
figure~\ref{phase1}.
Note that these spectra are obtained from the entire data set and not
from individual waves such as figures~\ref{phase1} and \ref{wiso}a,
thus providing conclusions on the average behaviour of sectors 1,2,3 and 4. These
conclusions are:

i)
figure~\ref{phafig1}b shows that
the dominant contribution of small scale energy comes from sector~2, that
is where the breaking of the wave occurs.

ii)
Sector 1 (figure~\ref{phafig1}a) shows higher levels of small scale
energy than sector 3 (figure~\ref{phafig1}c) this is consistent with
a major contribution from the parasitic capillaries developing in sector~1.

iii)
Sector 3 shows higher levels of large scale energy that sector~1,
this may be associated with the non-linearity of the waves in this part.

iv)
Sector 4 (figure~\ref{phafig1}d) shows a slightly higher level of
small scale energy that sector~3,
indicating some capillary leakage towards the downwind crest.

These conclusions support the view that the direction of propagation of waves
generated by breaking is from sector 2, to sector 1 to sector 4 of the subsequent
wave with no indication that these waves exist in sector 3.

\subsection{The distribution of capillary wave periods}
\label{capello}

%
\begin{figure}
\includegraphics[height=5.3cm]{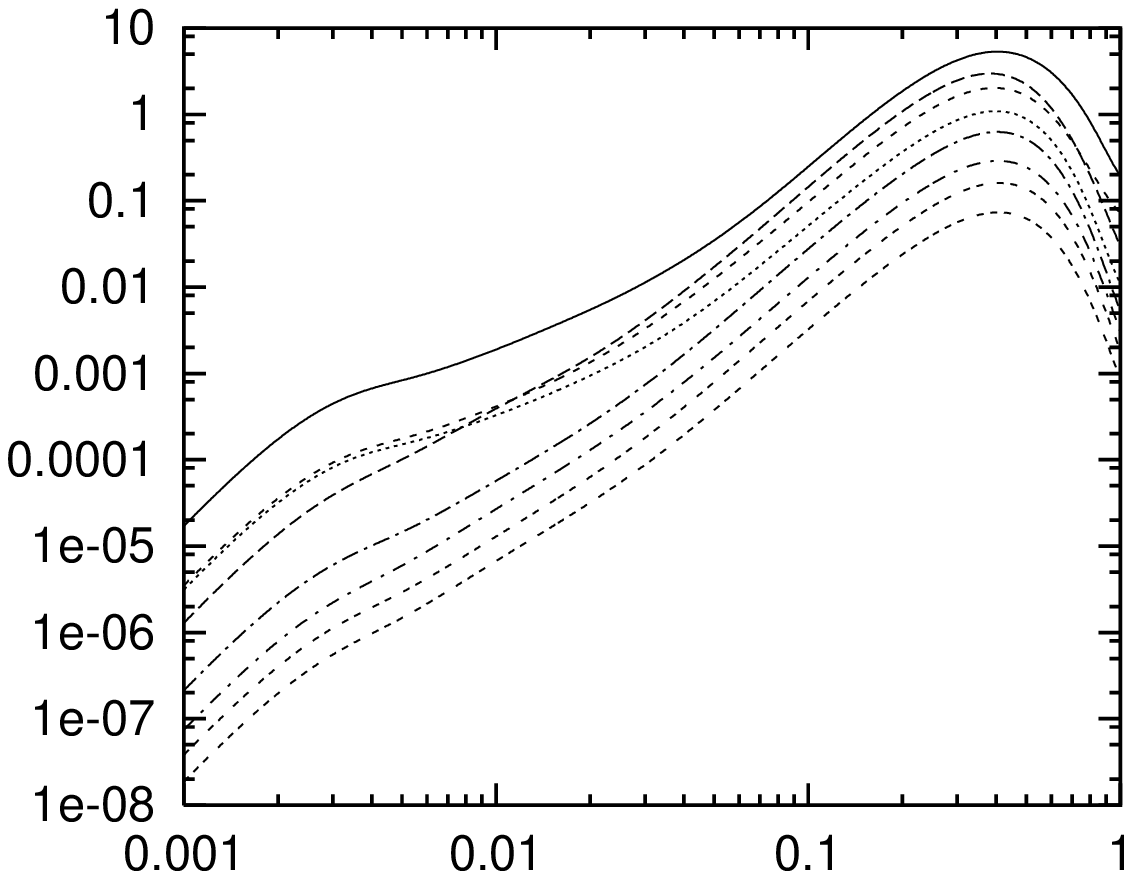}
\caption{Mexican-hat wavelet spectra for different length data
samples}
{From top to bottom $2^{10}$, $2^{11}$, $2^{12}$, $2^{13}$,
$2^{14}$, $2^{15}$, $2^{16}$, $2^{17}$ point long segments.}
\label{samplength}
\end{figure}

In the wavelet spectra shown in figure~\ref{phafig1}, samples larger
than $2^{17}$ have been used. One can see from these spectra
that the capillary wave period disappears from the total energy
spectrum based on $2^{17}$ points.
The sample length can be varied to get an idea of the
breadth of the distribution of capillary wave
periods at different scales.
Figure~\ref{samplength} shows wavelet spectra corresponding to
different sample sizes of the signal in figure~\ref{figu48f435}.
At small scales $0.001< \tau < 0.05$, the shape of the spectrum needs
samples longer than $2^{14}$ points to converge, whereas the spectrum
shape at large scales is not affected. This is an indication that the
distribution of capillary wave periods is broader at small scales than at
large ones.

\subsection{Large scale indentations:}
\label{indentations}

%
%
\begin{figure}
\includegraphics[height=4.7cm]{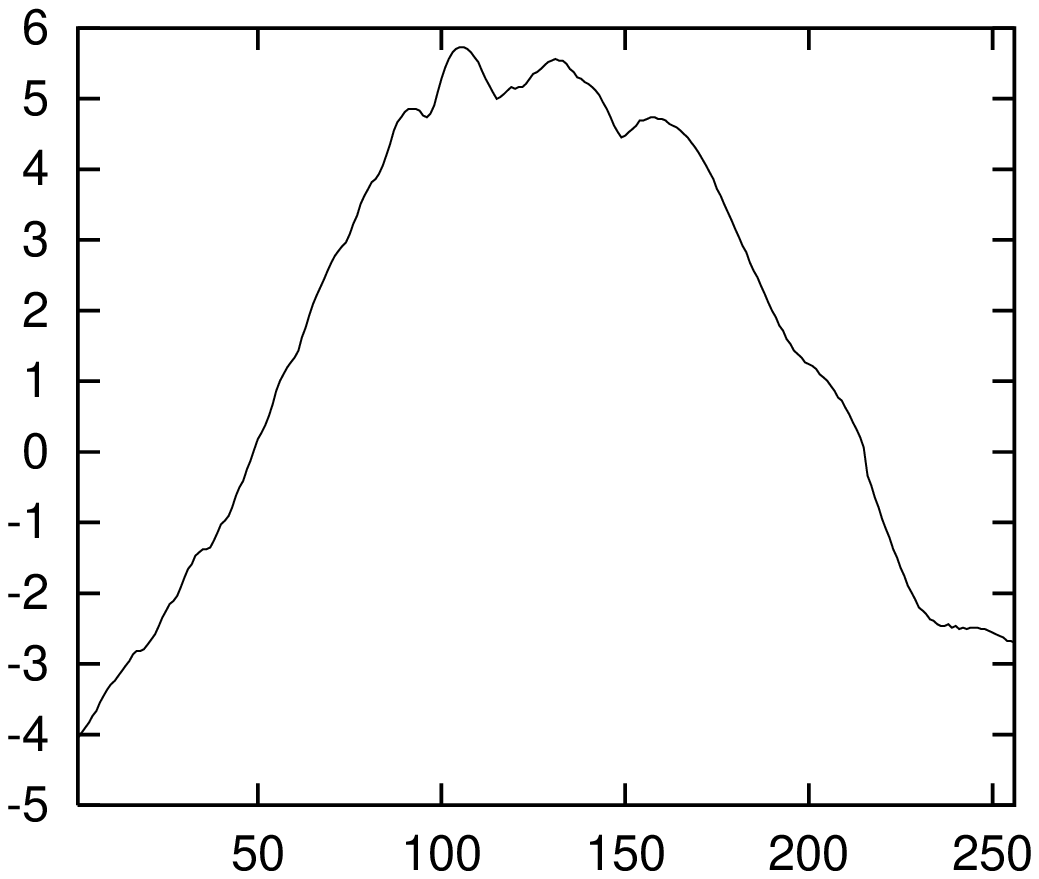}
\includegraphics[height=4.8cm]{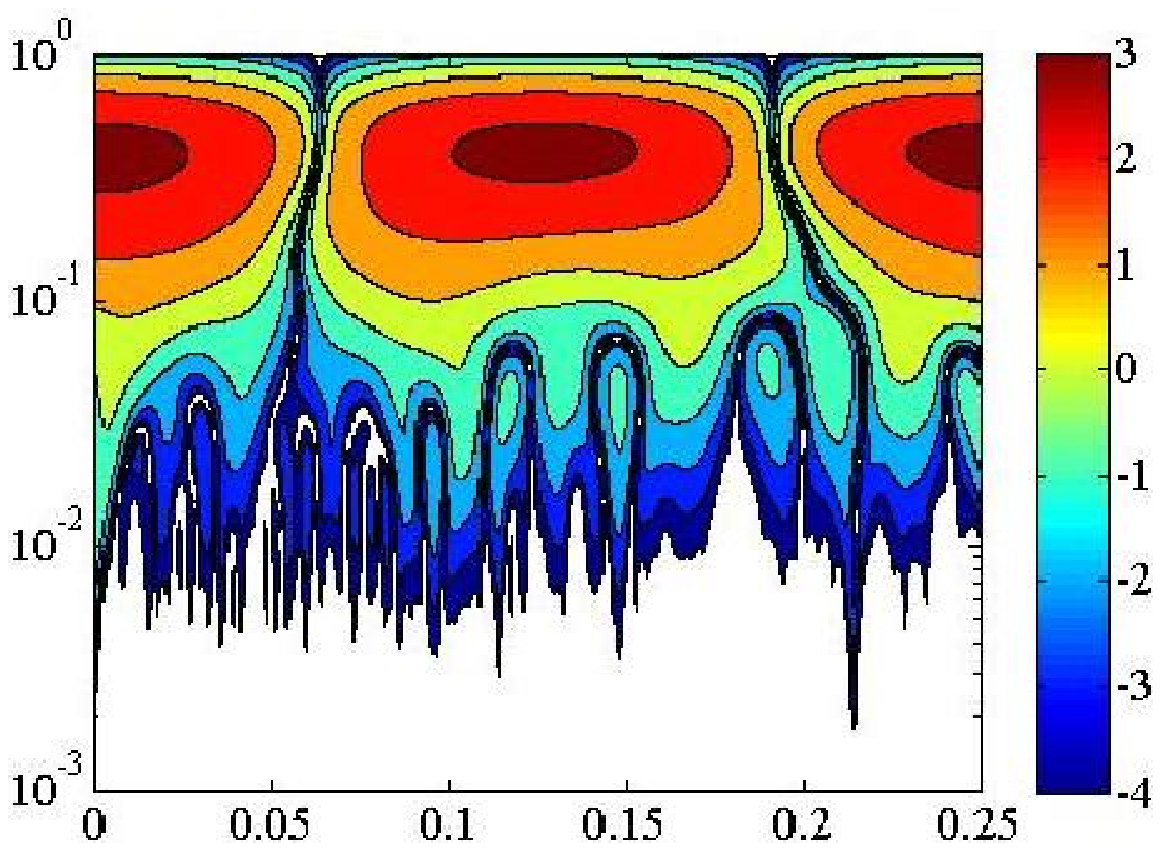}
\\
\hspace*{2cm} a) \hspace*{6.5cm} b)
\caption{Isolated wave with large indentations}
{
a) 256 points long signal in milliseconds,
b) wavelet transform of \ref{wiso2}a, iso-value sampling is
the same as in figure~\ref{wiso} and $t$ and $\tau$ are in seconds.}
\label{wiso2}
\end{figure}
\begin{figure}
\hspace{2cm}
\includegraphics[height=5cm]{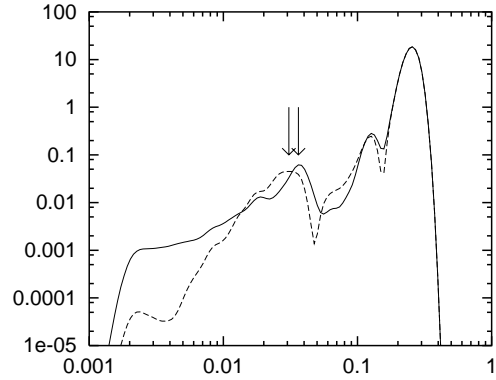}
\caption{Wavelet spectrum of the isolated wave in figure~\ref{wiso2}}
{Solid line wavelet spectrum of the entire wave,
dashed line wavelet spectrum conditioned on $h(t) > 5$.
Arrows indicate scales 0.031 and 0.036.}
\label{wisc2}
\end{figure}
%
%
We now focus on large scale indentations
like those in figure~\ref{wiso2}a. These are indentations
at scales larger than the capillary wave ones
 but still smaller than the large scale $\lambda$ educed
in \S 3(a).
Figure~\ref{wiso2}b is a plot of the wavelet transform of the signal.
A comparison of
figures~\ref{wiso}b and \ref{wiso2}b shows that the
wavelet-scales of the large scale indentations are
clearly an order of magnitude smaller than $\lambda$ and can be up to
an order of magnitude larger than the wavelet scales involved in capillary effects.
The scales associated with the large indentations in figure~\ref{wiso2}a
are not localised on the forward face and lie between t=0.031 s and
0.036 s. They are therefore clearly differentiated from the capillarity
waves analysed in the previous section.

Figure~\ref{wisc2} shows
the wavelet spectrum and the conditional wavelet
spectrum of signal in figure~\ref{wiso2}a.
The conditional spectrum is closer
to the $\tilde{E}(\tau) \sim \tau^3$ law indicating that, contrary to
capillary waves, the large indentations in figure~\ref{wiso2}a may be closer to
slope discontinuities,
but the range is too short to enable a definitive conclusion.
%

\section{Analysis of a fractal distribution of $\lambda$}
\label{sec7}

\begin{figure}
\includegraphics[width=12cm]{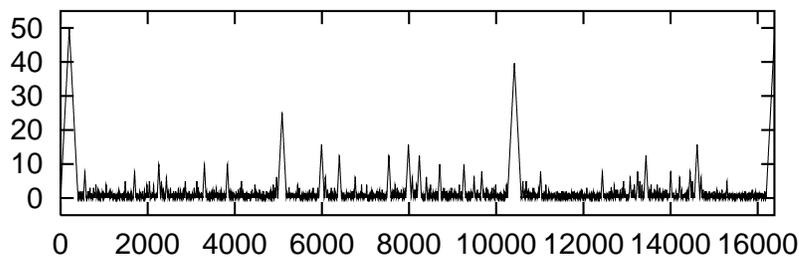}
\caption{
Signal $h(t)$ constructed as a sum of a
fractal distribution of $\Lambda$-crests for $D_1=0.5$, $\lambda_{min}=1$ and
$\lambda_{max}=100$.}
\label{figufra1.5}
\end{figure}

In this section we analyse a signal having a fractal distribution
of $\lambda$ (the distance between two consecutive zero-crossings), that is
[\cite{bv97}]:
\begin{equation}
M(\lambda) \, d\lambda = \frac{M_{max}}{ \lambda_{max}}
             \left ( \frac{\lambda_{max}}{\lambda} \right )^{D_1+1}
\, d \lambda
\label{fraclambda}
\end{equation}
where $M(\lambda)$ is the number density of $\Lambda$-crests of size
between $\lambda$ and $\lambda + d\lambda$
and we assume $0<D_1$ and $\lambda < \lambda_{max}$.
This signal is constructed as a sum of $\Lambda$-crests which are exactly
slope-discontinuities with the same angle between two slopes forming a
discontinuity. Hence the signal is constructed as a fractal distribution of
self-affine $\Lambda$-crests (see figure~\ref{figufra1.5}).
%
%

\cite{bv97} have shown that the Fourier spectrum of such a 1-D signal follows the power law
\begin{equation}
E(\omega) \sim
       \frac{1}{\omega^4}
       (\omega \lambda_{max})^{D_1}
\label{fracpow}
\end{equation}
for $\omega \lambda_{max} \gg 1$. Using (\ref{power}) we can deduce that
\begin{equation}
E(\tau) \sim \tau^{3-D_1}
\label{belpower}
\end{equation}
for $\frac{\tau}{\lambda_{max}} \ll 1$ and can also use wavelet spectra
conditioned on $h(t) > \epsilon$ to isolate $\Lambda$-crests and show that
for $\epsilon$ large enough
\begin{equation}
E_c(\tau) \sim \tau^{3}
\label{crestpower}
\end{equation}
even though $E(\tau) \sim \tau^{3-D_1}$
for $\frac{\tau}{\lambda_{max}} \ll 1$.
Hence, wavelet methods can be used to show the existence of
$\omega^{-4 + D_1}$ spectra but also to demonstrate,
when the case may be, that these power-law spectra are related to $\Lambda$-crests
which themselves have $\omega^{-4}$ energy spectra.

%
\begin{figure}
\includegraphics[height=4.5cm]{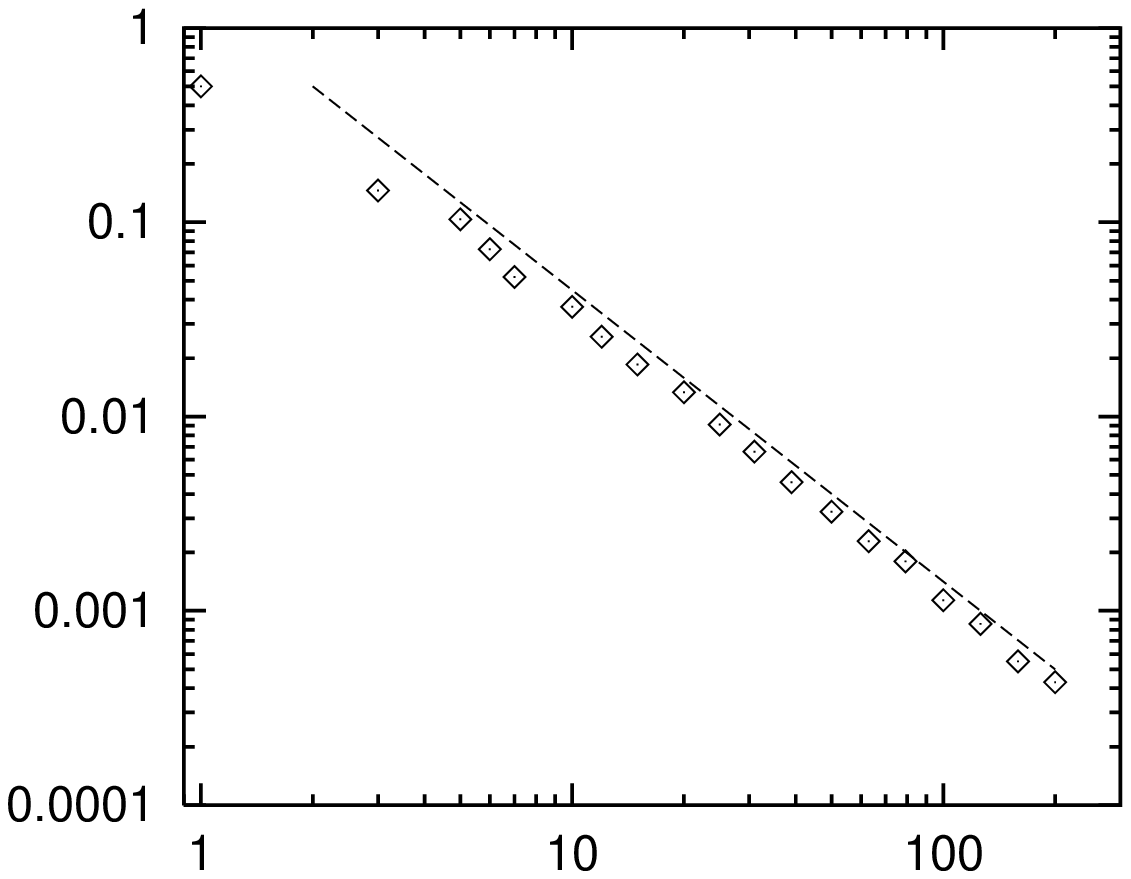}
\includegraphics[height=4.5cm]{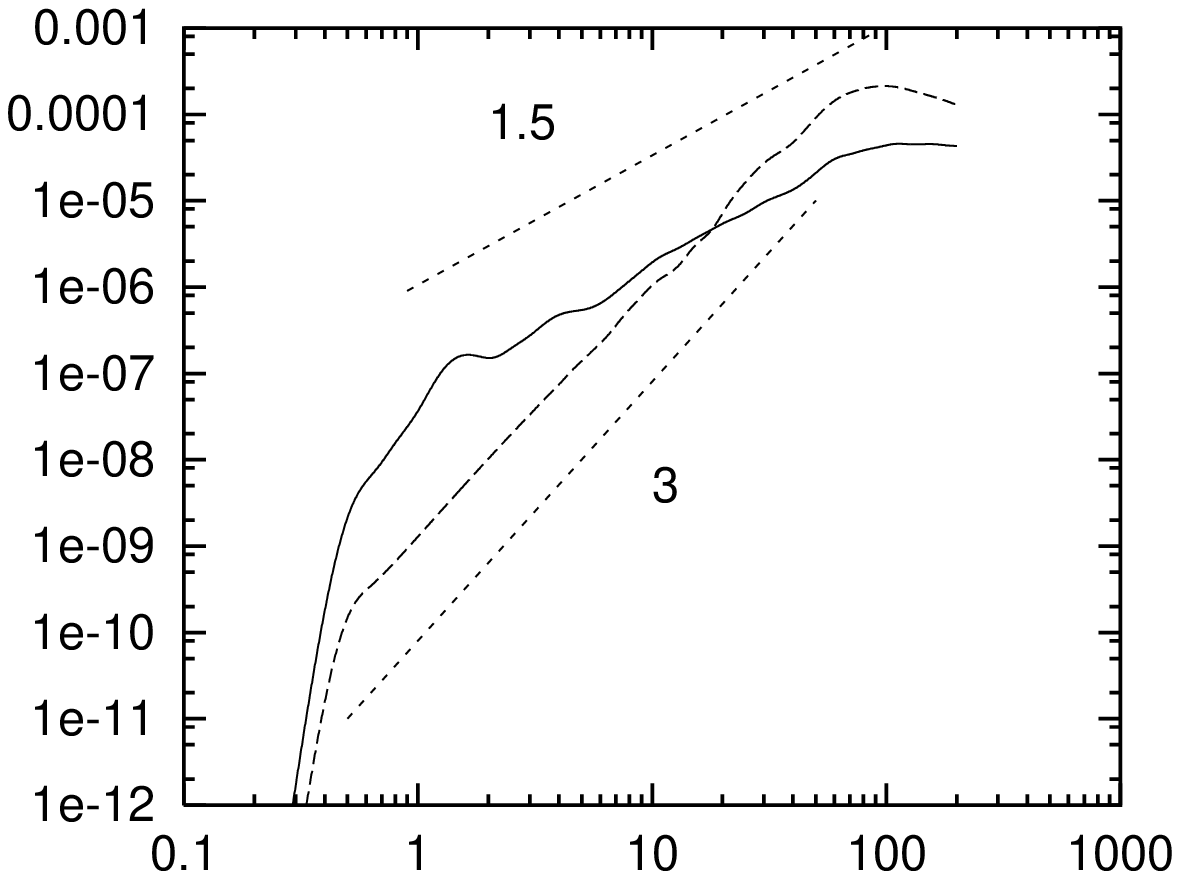}
\\[2ex]
\hspace*{2cm} a) \hspace*{6.cm} b)
\caption{Fractal and Wavelet analyses of signal shown in figure~\ref{figufra1.5}}
{a)Probability density function of $\lambda_{max}M(\lambda)$
as a function of $\lambda$
for the signal's first 300000 points.
The straight line is the slope -1.5.
b) Wavelet transform of the signal's first 131072 points based on the
Morlet wavelet. The solid curve corresponds to the entire sample,
the dash curve to the
condition $\epsilon > 10$.}
\label{lambdafra1.5}
\end{figure}
%
%

Figure~\ref{lambdafra1.5}a is a log-log plot of $\lambda_{max}M(\lambda)$
versus $\lambda$
enabling direct verification of the fractal distribution
of $\lambda$ with $D_1 = 0.5$.
Figure~\ref{lambdafra1.5}b is a plot of
the wavelet spectrum of the signal in figure~\ref{figufra1.5}
(using the 131072 points of this signal) and
it is based on the Morlet wavelet. The solid
line corresponds to the entire signal, the dash line to the signal
conditioned on $h(t)>\epsilon > 10$.
We can draw the following conclusions:

i)
there is no particular peak in these spectra as was the case in the
experimental data studied in the previous sections.
This is consistent with the notion of a fractal distribution
which implies that there is no privileged scale.

ii)
The spectrum of the entire data is dominated by the fractal distribution
of the peaks: the law for the wavelet-spectrum is
$\tilde{E}(\tau) \sim \tau^{1.5}$
which yields a Fourier spectrum $E(\omega) \sim \omega^{-2.5}$
in agreement with
relations (\ref{fracpow}) and (\ref{belpower}) for $D_1 = 0.5$.

iii)
The spectrum associated with the $\Lambda$-crests is educed by the conditioned
spectrum based on the condition $h(t) > 10$.
We find $\tilde{E}_c(\tau) \sim \tau^{3}$
which agrees with relation~(\ref{crestpower}) for slope discontinuities.
\\[2ex]
For the sake of completeness we should also mention the possibility of
a fractal distribution of scales over a limited range which would generate a scale-dependent behaviour of the kind of the `parabolic scale invariance' introduced in [\cite{Queiros-Conde-2003}].

%
\section{Conclusion}
In this paper we have used conditional spectra based on wavelet
decompositions to analyse time series of displacement of tank waves.
We have shown that wavelets can educe capillary waves from the signal.
Wavelet spectrum analysis has also enabled us to quantify the relative contribution of
peaks and troughs to the energy spectrum.
Wave peaks are close to slope discontinuities whereas wave troughs are not.
Wavelet spectra conditioned on wave sectors
show that capillary waves are mainly located in
sector 1 with some capillary leakage downwind from the peak of the wave
indicating that capillary waves
propagate downwind.
\\[2ex]
The wavelet analysis we presented here was tuned to our particular application. There are many examples of forced microscale waves,
for example the wave displacement in driven metal plates in [\cite{Miquel-Mordant-2011}]; though the turbulence there is classified as weak it is richer in scales than the case presented in this paper. Such cases would be interesting study cases as preliminary to the use of wavelet analysis to the study of fully developed turbulence.

\section*{Acknowledgements}
We are grateful to W. L. Peirson for enabling us to use his data.
JCV acknowledges supports from the Royal Society.
%


\appendix

\section{Mexican-hat wavelet}
\label{apmex}

Wavelets based on the Gaussian function have the form:
\begin{equation}
\psi(t) = \frac{d^n}{d t^n} e^{-\frac{1}{2} t^2},
\end{equation}
the case $n=2$ corresponds to the Mexican-hat wavelet. Their Fourier
transforms have the simple form:
\begin{figure}
\includegraphics[width=5cm,angle=-90]{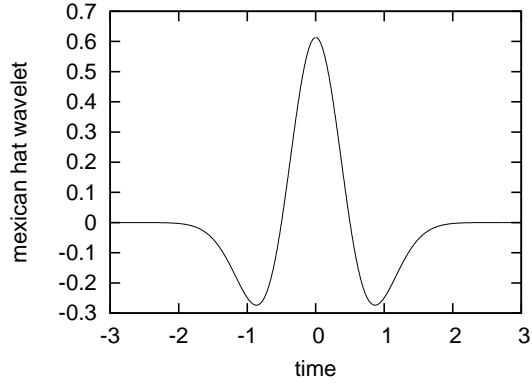}
\caption{\label{fig18}Typical Mexican-hat wavelet.}
\end{figure}
\begin{equation}
\hat{\psi}(\omega) = (i\omega)^n e^{-\frac{1}{2} \omega^2}.
\label{eqft}
\end{equation}
These wavelets are interesting for the study of the zero-crossings
of signals but can be misleading in the study of spectrum power laws.
Indeed, with these functions as mother wavelets,
the wavelet spectrum asymptotic limit is
\[
\lim_{\tau \to 0} \tilde{E}(\tau) \sim \tau^{2n},
\]
whatever the Fourier spectrum.
This is due to the fact that in practise a power law spectrum
is not verified over an infinite range of scale but has an upper
$\omega_{max}$ and lower $\omega_{min}$ cut-off scale.
Using this remark and plugging (\ref{eqft}) into the definition of
the wavelet spectrum (\ref{eqspew}), it yields:
\[
\tilde{E}(\tau) = \int_{\omega_{min}}^{\omega_{max}}
               \left | \hat{h}(\omega) \right |^{2}
               (\tau \omega)^{2n} e^{-(\tau \omega)^2} \, d \omega,
\]
that is
\[
\tilde{E}(\tau) = \tau^{2n} \int_{\omega_{min}}^{\omega_{max}}
               \left | \hat{h}(\omega) \right |^{2}
                       \omega^{2n} e^{-(\tau \omega)^2} \, d \omega.
\]
Because $\omega_{min}$ and $\omega_{max}$ are finite $\tau$
can go to 0 while $\omega$ is bounded, and
\[
\lim_{\tau \to 0} e^{-(\tau \omega)^2} = 1
\]
is valid outside the range
$\frac{1}{\omega_{max}} < \tau < \frac{1}{\omega_{min}}$.
Then
\[
\tilde{E}(\tau) \sim \tau^{2n} \int_{\omega_{min}}^{\omega_{max}}
                         \left | \hat{h}(\omega) \right |^{2}
                         \omega^{2n} \, d \omega,
\]
that is
\[
\tilde{E}(\tau) \sim \tau^{2n}
\]
whatever the form of $\left | \hat{h}(\omega) \right |^{2}$. This asymptotic
behaviour proper to the wavelet can parasite the spectrum power law we are
looking for. Figure~\ref{speclocf} shows the wavelet spectrum form 3 wavelets:
$\frac{d^2}{d t^2} e^{-\frac{1}{2} t^2}$,
$\frac{d^4}{d t^4} e^{-\frac{1}{2} t^2}$ and the Morlet wavelet
(see \S~\ref{apmor}).
\begin{figure}
\includegraphics[height=5cm]{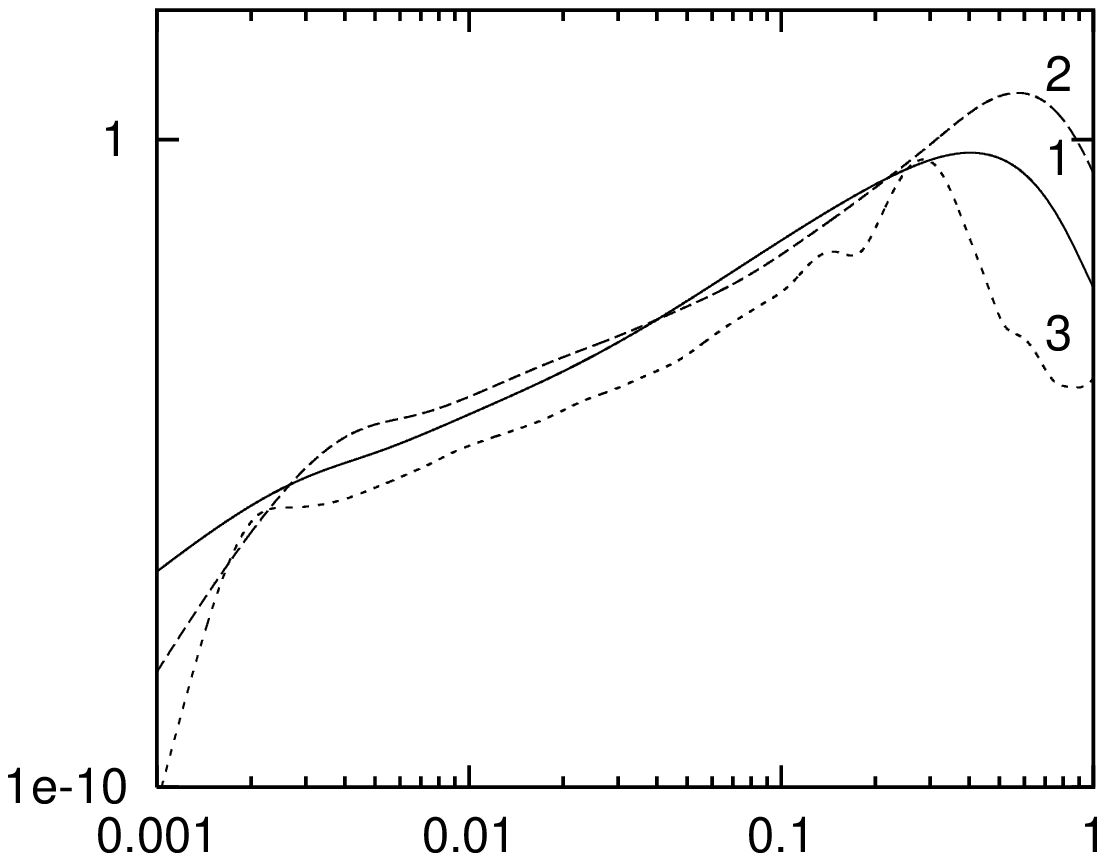}
\caption{$\tilde{E}(\tau)$ as a function of $\tau$ for 3 different wavelets}
{
1) $\frac{d^2}{d t^2} e^{-\frac{1}{2} t^2}$,
2) $\frac{d^4}{d t^4} e^{-\frac{1}{2} t^2}$,
3) Morlet wavelet.
the cut-off is the clearer with the Morlet wavelet.}
\label{speclocf}
\end{figure}
%


\section{Morlet wavelet}
\label{apmor}

The Morlet wavelet is defined as:
\begin{equation}
\psi(t) = e^{-\frac{1}{2}t^2} e^{it}.
\label{wmorl}
\end{equation}
and its Fourier transform is
%
$
\hat{\psi}(\omega) = e^{-\frac{1}{2}(\omega-\omega_0)^2}.
$
If we use this wavelet in the computation of (\ref{eqspew}) then
\begin{equation}
\tilde{E}(\tau) \sim \int_{\omega_{min}}^{\omega_{max}}
                         \left | \hat{h}(\omega) \right |^{2}\, d\omega
\end{equation}
when $\tau \to 0$ and there is
no asymptotic power law introduced in that limit.
when $\tau \to 0$.
On the other hand due to its cosine-like form, Morlet wavelet tends to
focus on a scale when it is periodic at the expense of the resolution
of the power spectrum (where it exists). As it appears in figure~\ref{speclocf}
apart from the determination of cut-off scales, the Mexican-hat wavelet is
more appropriate for the determination of the power law of a spectrum.
%


%


%

\section*{\it Journal reference:}




\end{document}